\documentclass[a4paper,12pt]{article}
\pdfoutput=1

\usepackage[utf8]{inputenc}
\usepackage{jheppub}
\usepackage{amsmath}
\usepackage{bbm}
\usepackage{graphicx}
\usepackage{amssymb}
\usepackage{booktabs}
\usepackage{caption}
\usepackage{subcaption}
\usepackage{slashed}
\usepackage[textwidth=22mm]{todonotes}

	\newcommand{\Tr}{\mathrm{Tr}}

	\newcommand{\vtr}{\Phi_{\mathrm{tr}}^{(1)}}
	\newcommand{\phin}{\underline{\phi_0}}
	\newcommand{\phio}{\underline{\phi_1}}

	\newcommand{\sn}{\mathrm{n}}	
	\newcommand{\sk}{\mathrm{k}}
	
		\newcommand{\sdot}{\circ}

\subheader{\hfill \begin{tabular}{r} \texttt{QMUL-PH-18-22} \end{tabular}}

\title{Spinor-helicity and the algebraic classification of higher-dimensional spacetimes}
\author[a]{Ricardo Monteiro,}
\author[b]{Isobel Nicholson,}
\author[b]{and Donal O'Connell}
\affiliation[a]{Centre for Research in String Theory, School of Physics and Astronomy, Queen Mary University of London, 327 Mile End Road, London E1 4NS, UK}

\affiliation[b]{
Higgs Centre for Theoretical Physics, School of Physics and Astronomy, The University of Edinburgh, Edinburgh EH9 3JZ, Scotland, UK%
}
\emailAdd{ricardo.monteiro@qmul.ac.uk,i.nicholson@sms.ed.ac.uk,\\donal@ph.ed.ac.uk}

\abstract{
The spinor-helicity formalism is an essential technique of the amplitudes community. We draw on this method to construct a scheme for classifying higher-dimensional spacetimes in the style of the four-dimensional Petrov classification and the Newman-Penrose formalism. We focus on the five-dimensional case for concreteness. Our spinorial scheme naturally reproduces the full structure previously seen in both the CMPP and de Smet classifications, and resolves longstanding questions concerning the relationship between the two classifications.
}

\begin{document}
\maketitle

\section{Introduction}

Representations of the Lorentz group play a prominent role in particle physics. Particle states are famously classified according to irreducible representations, and the requirement of Lorentz invariance strongly constrains their interactions. This constraint is particularly powerful when dealing with massless particles.
In four spacetime dimensions, the isomorphism $SO(3,1)\cong SL(2,\mathbb{C})/\mathbb{Z}_2$ allows us to write any massless momentum as a product of two spinors, $k_\mu\mapsto\lambda_\alpha \tilde \lambda_{\dot \alpha}$ \cite{Wigner:1939cj}. For the scattering of massless particles, an S-matrix element is a function of these spinors only, and the helicities $h_i$ of each particle fix the relative homogeneity weight of the function for each type of spinor. This is known as the spinor-helicity formalism, and it has become a major tool in high-energy physics. See, e.g., ref.~\cite{Elvang:2013cua} for a recent review of this formalism and its applications.

General relativity has also seen fruitful applications of this type of idea, starting with Penrose's spinorial approach \cite{Penrose:1960eq} and its development into the Newman-Penrose formalism \cite{Newman1962}. The basic principles are to define a frame $e^\mu_{\;\;M}$ that takes us from coordinate space to the tangent space, $\eta_{MN}=g_{\mu\nu}\,e^\mu_{\;\;M}\,e^\nu_{\;\;N}$, and then to explore the isomorphism $SO(3,1)\cong SL(2,\mathbb{C})/\mathbb{Z}_2$ for the tangent space Lorentz transformations. For instance, the Weyl tensor $C_{\mu\nu\rho\sigma}$ is described in tangent space by a rank 4 spinor $\psi_{\alpha\beta\gamma\delta}$ and its complex conjugate. The algebraic classification of this rank 4 spinor elegantly reproduces the Petrov classification of four-dimensional spacetimes \cite{Petrov}, which had a profound impact in the development of general relativity; see, e.g., refs.~\cite{Stephani:2003tm,Griffiths:2009dfa}. In particular, the Kerr solution, which represents a vacuum asymptotically flat stationary black hole, and is perhaps the most important exact solution of astrophysical interest, was originally discovered by imposing a condition of algebraic specialty \cite{Kerr:1963ud}.

There are a variety of motivations for extending these constructions to higher spacetime dimensions. In the case of general relativity, extra dimensions are naturally motivated by string theory, and also by the fact that the number of spacetime dimensions is the natural parameter of the vacuum Einstein equations. Indeed, the catalogue of higher-dimensional vacuum asymptotically flat black hole solutions is incredibly rich, in contrast with the four-dimensional case, where the unique solution is the Kerr black hole; see, e.g.,~ref.~\cite{Emparan:2008eg,Emparan:2009vd,Horowitz:2012nnc,Dias:2015nua} for reviews.

In the case of particle physics, analogous motivations apply to developing the spinor-helicity formalism in various dimensions. There is also a more practical application to the computation of S-matrix elements in dimensional regularisation, where the loop momenta cannot be restricted to four dimensions. An elegant extension of the spinor-helicity formalism approach to higher dimensions was presented in~\cite{Cheung2009}, where the main focus was on six dimensions. The method was extended to general dimensions in~\cite{CaronHuot:2010rj,Boels:2012ie}. In our paper, we will apply this extension to the algebraic classification of solutions in general relativity.

As we mentioned, the space of solutions to the vacuum Einstein equations in higher dimensions is much richer than that in four dimensions, and the question of extending the Petrov classification naturally arose in the past. In fact, different approaches have been taken. Coley, Milson, Pravda and Pravdova (CMPP) defined a classification~\cite{Coley2004a, Coley2004b} that has been investigated over many years, for example in~\cite{Coley2004c, Pravda2005, Pravda2007, Pravdova2008, Coley2009, Ortaggio2012, Coley2012,Ortaggio:2017ydo}; see \cite{Ortaggio:2012jd} for a review. In analogy to the four-dimensional story, the classification is based on the grouping of Weyl tensor components according to boost weight. Subgroups within the groups of boost-weighted components were found by Coley and Hervik in \cite{Coley2009}, and in \cite{Coley2012} these sub-types were investigated in five dimensions. The CMPP classification has not been studied from a purely spinorial approach.

A different classification had been previously constructed by de Smet~\cite{deSmet2002} for five-dimensional spacetimes, based on the factorisation properties of the Weyl spinor. This spinorial approach can also be considered a natural extension of the four-dimensional story, and yet it takes a very different form to the CMPP construction. An in-depth comparison by Godazgar~\cite{Godazgar2010} showed that there was poor agreement in what was considered algebraically special by the de Smet classification versus the CMPP classification. None of two appeared to be the `finest' classification, since a solution could be special in one classification and general in another. 

There are two main goals to our paper. The first is to apply the higher-dimensional spinor-helicity formalism of ref.~\cite{Cheung2009} to the algebraic classification of solutions of the Einstein equations, in the spirit of the spinorial approach of Penrose. The second is to show the versatility of this spinorial approach, which exhibits manifestly the two relevant types of spinor spaces, by clarifying the relation between the CMPP and the de Smet classifications, and the question of the `finest' algebraic classification. We will be mostly interested in five-dimensional solutions, where the spinorial formalism is based on the isomorphism $SO(4,1)\cong Sp^\ast(1,1)/\mathbb{Z}_2$, but we will also briefly discuss the six-dimensional case in order to demonstrate generic features. We will be careful to describe when we consider reality conditions in our spinorial formalism, so that it can be applied both to real spacetimes and to potentially interesting cases of complexified spacetimes.

In addition to the classification of the Weyl tensor, we will study -- for illustration and as customary in this context -- the classification of its analogue in electromagnetism, the Maxwell field strength. There is a modern motivation to include this. A relation between gravity and gauge theory known as the `double copy' has emerged from the study of scattering amplitudes in quantum field theory \cite{Bern:2008qj,Bern:2010ue}. This relation, which applies in any number of spacetime dimensions, has a counterpart in terms of solutions to the field equations. It can be expressed most clearly for certain algebraically special solutions, namely Kerr-Schild spacetimes \cite{Monteiro:2014cda,Luna:2015paa,Ridgway:2015fdl,Luna:2016due,Bahjat-Abbas:2017htu,Carrillo-Gonzalez:2017iyj}, but it should apply more generally \cite{Monteiro:2011pc,Saotome:2012vy,Anastasiou:2014qba,Cardoso:2016ngt,Goldberger:2016iau,Luna:2016hge,Cardoso:2016amd,Goldberger:2017frp,Adamo:2017nia,Luna:2017dtq,Goldberger:2017vcg,Chester:2017vcz,Goldberger:2017ogt,Li:2018qap,LopesCardoso:2018xes,Ilderton:2018lsf,Shen:2018ebu,Anastasiou:2018rdx,Lee:2018gxc,Plefka:2018dpa}. It is clear from these developments that there is a close relation between the algebraic properties of spacetimes and those of gauge field configurations. Indeed, it will be obvious from our results that an analogy exists. We hope to address elsewhere how this analogy can be turned into a precise double-copy relationship. 

This paper is organised as follows. In Section~\ref{recap}, we review the four-dimensional spinorial approach to the Petrov classification. We introduce in Section~\ref{spinors} the five-dimensional spinorial formalism. The five-dimensional algebraic classification is described in Section~\ref{maxwell} for the field strength tensor, for illustration, and then in Section~\ref{GR} for the Weyl tensor. The extension of this spinorial approach to higher dimensions is discussed in Section~\ref{highdim}. We conclude with a discussion of the results and possible future directions in Section~\ref{conclusion}.

\section{Review of the four-dimensional story}\label{recap}

In this section, we begin by discussing the familiar case of spinors in four dimensions to set up our notation. We then review the Petrov classification for four-dimensional spacetimes. This classification can be understood from a variety of perspectives; we emphasise the Newman-Penrose (NP) approach~\cite{Penrose1984, Newman1962} because it is closest in spirit to our approach in five dimensions.

\subsection{Spinors in four dimensions}
\label{sec:4dspinors}

In flat Minkowski space, the Clifford algebra is 
\begin{equation}
	\sigma^\mu{}_{\alpha\dot{\alpha}}\,\tilde{\sigma}^\nu{}^{\dot{\alpha}\beta} + \sigma^\nu{}_{\alpha\dot{\alpha}}\,\tilde{\sigma}^\mu{}^{\dot{\alpha}\beta}=-2 \eta^{\mu\nu} \,\mathbbm{1}_\alpha{}^\beta,
\end{equation}
where $\eta_{\mu\nu}$ is the Minkowski metric.\footnote{We work in the mostly-plus signature $(-,+,+,\cdots,+)$ in both four and higher dimensions.}
To be explicit, we choose a basis of $\sigma^\mu$ matrices given by
\begin{align}
\sigma^0 = \begin{pmatrix} 1 & 0 \\ 0 & 1 \end{pmatrix}, \quad
\sigma^1 = \begin{pmatrix} 0 & 1 \\ 1 & 0 \end{pmatrix}, \quad 
 \sigma^2 = \begin{pmatrix} 0 & -i \\ i  & 0 \end{pmatrix}, \quad
\sigma^3 = \begin{pmatrix} 1 & 0 \\ 0 & -1 \end{pmatrix},
\label{eq:sigma4def}
\end{align}
while the $\tilde \sigma^\mu$ matrices are
\begin{align}
\tilde \sigma^0 = \begin{pmatrix} 1 & 0 \\ 0 & 1 \end{pmatrix}, \quad 
\tilde \sigma^1 = -\begin{pmatrix} 0 & 1 \\ 1 & 0 \end{pmatrix}, \quad 
\tilde \sigma^2 = -\begin{pmatrix} 0 & -i \\ i  & 0 \end{pmatrix},  \quad
\tilde \sigma^3 = -\begin{pmatrix} 1 & 0 \\ 0 & -1 \end{pmatrix}.
\label{eq:tsigma4def}
\end{align}

For any non-vanishing null vector $V$, the matrices $V \cdot \sigma$ and $V \cdot \tilde \sigma$ have rank 1. Hence we may construct solutions of the (massless) Dirac equations:
\begin{align}
V &\cdot \sigma_{\alpha \dot \alpha}\, \tilde \lambda^{\dot \alpha} =0, \\
V &\cdot \tilde{\sigma}^{\dot{\alpha}\alpha}\, \lambda_\alpha=0.
\end{align} 
These spinors can be normalised so that $V \cdot \sigma_{\alpha \dot \alpha} =-\sqrt{2}\,  \lambda_\alpha \tilde \lambda_{\dot \alpha}$. 
We may raise and lower the indices $\alpha$ and $\dot \alpha$ on these spinors with the help of the two-dimensional Levi-Civita tensor. We choose conventions such that $\epsilon^{12}=1$, $\epsilon_{12}=-1$ and $s^\alpha=\epsilon^{\alpha \beta} \,s_\beta$ while $\tilde{s}{}^{\dot{\alpha}} = \epsilon^{\dot{\alpha}\dot{\beta}}\, \tilde {s}_{\dot{\beta}}.$

In the curved space case, we simply introduce a frame $e^\mu{}_M$, such that
\begin{equation}
g^{\mu\nu} = e^\mu{}_M\,e^\nu{}_N\, \eta^{MN} .
\end{equation}
On the tangent space at each point, the Clifford algebra can be written as before,
\begin{equation}
	\sigma^M{}_{\alpha\dot{\alpha}}\,\tilde{\sigma}^N{}^{\dot{\alpha}\beta} + \sigma^N{}_{\alpha\dot{\alpha}}\,\tilde{\sigma}^M{}^{\dot{\alpha}\beta}=-2 \eta^{MN} \,\mathbbm{1}_\alpha{}^\beta,
\end{equation}
whereas
\begin{equation}
	\sigma^\mu{}_{\alpha\dot{\alpha}}\,\tilde{\sigma}^\nu{}^{\dot{\alpha}\beta} + \sigma^\nu{}_{\alpha\dot{\alpha}}\,\tilde{\sigma}^\mu{}^{\dot{\alpha}\beta}=-2 g^{\mu\nu} \,\mathbbm{1}_\alpha{}^\beta,
\end{equation}
with $\sigma^\mu = e^\mu{}_M \sigma^M$, and a similar definition for $\tilde \sigma$.
We use the explicit Clifford bases of equations~\eqref{eq:sigma4def} and ~\eqref{eq:tsigma4def} in the tangent space.

It may be worth commenting briefly on reality conditions in four dimensions, since the reality conditions in five dimensions will play a more significant role later. The Lorentz group in real Minkowski space is $SL(2, \mathbb{C})/\mathbb{Z}_2$. It is consistent to choose a basis of Hermitian $\sigma$ matrices -- and indeed we have chosen such a basis in equations~\eqref{eq:sigma4def} and~\eqref{eq:tsigma4def}. Then, given a real null vector $V$, we may choose our spinors $\lambda$ and $\tilde \lambda$ such that $\lambda^\dagger = \tilde \lambda$. This is consistent with the choice that $V \cdot \sigma_{\alpha \dot \alpha} = -\sqrt{2}\, \lambda_\alpha \tilde \lambda_{\dot \alpha}$.

\subsection{The four-dimensional Newman-Penrose tetrad}
\label{sec:4dNP}

In four dimensions, the NP formalism~\cite{Penrose1984, Newman1962} exploits the correspondence between the Lie algebras $\mathfrak{so}(4)$ and $\mathfrak{su}(2)\times \mathfrak{su}(2)$. A key element of the method is the spinorial construction of a particular basis set of vectors, known as the NP tetrad. We begin by choosing two null vectors $k^\mu$ and $n^\mu$ which satisfy $k\cdot n\neq 0$, and constructing an associated basis of spinors $\{o_\alpha,\imath_\alpha\}$ by solving the equations
		\begin{equation}\label{dirac}
			k\cdot \tilde{\sigma}_{\dot{\alpha}\alpha}\, o^\alpha=0,\hspace{10pt}n\cdot \tilde{\sigma}_{\dot{\alpha}\alpha}\,\imath^\alpha=0.
		\end{equation} 
Since $k\cdot n\neq 0$, we may normalise the vectors so that $k \cdot n = -1$, and also normalise our spinors so that $o^\alpha \,\imath_\alpha=1$.  

Similarly, we construct a conjugate basis by solving the equations
\begin{equation}\label{tdirac}
		k\cdot{\sigma}_{\alpha\dot{\alpha}}\, \tilde {o}^{\dot{\alpha}}=0,\hspace{10pt}n\cdot {\sigma}_{\alpha\dot{\alpha}}\,\tilde {\imath}^{\dot{\alpha}}=0,
\end{equation}
to find the dual spinors 
$\{\tilde {o}_{\dot{\alpha}}, \tilde {\imath}_{\dot \alpha}\}$, which we also normalise so that $\tilde {o}^{\dot{\alpha}} \tilde {\imath}_{\dot{\alpha}}=1$. For real $k$ and $n$, we may take $\tilde o = o^\dagger$ and $\tilde \imath = \imath^\dagger$ as discussed in section~\ref{sec:4dspinors}.

Let us now complete the construction of the NP tetrad of vectors using our spinor basis. The tetrad includes the vectors $k$ and $n$, so we must find two more. Since the spinor basis is complete, we can construct the last two elements of the NP tetrad, $m$ and $\widetilde{m}$, from
	\begin{equation}
		m^\mu=\frac{1}{\sqrt 2}\, \sigma^\mu_{\alpha\dot{\alpha}}\,\imath^\alpha \,\tilde{o}^{\dot{\alpha}},
		\qquad \widetilde{m}^\mu=\frac{1}{\sqrt 2}\, \sigma^\mu_{\alpha\dot{\alpha}}\,o^\alpha \,\tilde{\imath}^{\dot{\alpha}}.
	\end{equation}  
Of course, when $k$ and $n$ are real, $\widetilde{m}$ is the conjugate of $m$. It is then a straightforward exercise to show that all four vectors in the NP tetrad are null, and satisfy $-k\cdot n=m\cdot\widetilde{m}=1$ with all other dot products vanishing. Furthermore, by use of these properties the spinorial completeness relation transmutes into the NP metric,
		\begin{equation}
			g^{\mu\nu}=-k^\mu\,  n^\nu-k^\nu\,  n^\mu+m^\mu\,  \widetilde{m}^\nu+m^\nu \, \widetilde{m}^\mu.
		\end{equation}
Thus we can fully describe the spacetime in terms of spinors.

	\subsection{The Petrov classification for 2-forms and the Weyl spinor}

These four-dimensional spinors make it possible to rewrite the field strength 2-form and the Weyl tensor in a convenient form. For an arbitrary 2-form $F_{\mu\nu}$, we can build a complex symmetric spinor
\begin{equation}
	\Phi_{\alpha \beta}=F_{\mu\nu}\, \sigma^{\mu\nu}{}_{\alpha \beta},
\end{equation}
where $\sigma^{\mu\nu}{}_{\alpha \beta}=\frac{1}{2}\left(\sigma^\mu{}_{\alpha\dot{\gamma}}\, \tilde{\sigma}^{\nu\dot{\gamma}}{}_\beta-\sigma^\nu{}_{\alpha\dot{\gamma}}\, \tilde{\sigma}^{\mu\dot{\gamma}}{}_\beta\right)$. The symmetric two-dimensional matrix $\Phi_{\alpha\beta}$ is parameterised by three complex scalars,
\begin{equation}\label{4D2formdecomposition}
	\phi_0=	\Phi_{\alpha \beta}\, o^\alpha \,o^\beta,\hspace{10pt}	\phi_1=	\Phi_{\alpha \beta}\, o^\alpha \,\imath^\beta,\hspace{10pt}	\phi_2=	\Phi_{\alpha\beta} \,\imath^\alpha\, \imath^\beta.
\end{equation}
Similarly, we can build a symmetric 4-spinor, known as the Weyl spinor, from the Weyl tensor $C_{\mu\nu\rho\sigma}$
\begin{equation}
\Psi_{\alpha \beta\gamma\delta}=C_{\mu\nu\rho\sigma}\,\sigma^{\mu\nu}{}_{\alpha \beta}\,\sigma^{\rho\sigma}{}_{\gamma\delta}.
\end{equation}
The Weyl spinor can be decomposed into 5 complex scalars defined by:
 \begin{equation}\label{4DGRdecomposition}
 \begin{split}
\psi_0=&	\Psi_{\alpha \beta\gamma\delta}\, o^\alpha \,o^\beta o^\gamma o^\delta,\hspace{10pt}
\psi_1=	\Psi_{\alpha \beta\gamma\delta}\, o^\alpha\, o^\beta o^\gamma\, \imath^\delta,\hspace{10pt}
\psi_2=	\Psi_{\alpha \beta\gamma\delta} \,o^\alpha \,o^\beta \,\imath^\gamma\, \imath^\delta,
\\\psi_3=&	\Psi_{\alpha \beta\gamma\delta} \,o^\alpha\, \imath^\beta\, \imath^\gamma\, \imath^\delta,\hspace{10pt}
\psi_4=	\Psi_{\alpha \beta\gamma\delta}\, \imath^\alpha \, \imath^\beta\, \imath^\gamma\, \imath^\delta.
 \end{split}
\end{equation}

The Petrov classification~\cite{Petrov2000} is a way of categorizing Weyl and field strength spinors depending on how ``algebraically special'' they are. It is well known that a symmetric $SU(2)$ $n$-spinor will always factorise into the symmetrisation of $n$ basic spinors. 
The idea of the Petrov classification is that the more of these individual spinors that are the same (up to scale), the more special the original $n$-spinor is. For example, a field strength spinor $\Phi_{\alpha \beta}=\alpha_{(\alpha}\beta_{\beta)}$ is algebraically special if and only if $\beta \propto \alpha$. This also has an interpretation in terms of the complex scalars $\phi_i$ (and $\psi_i$ for the Weyl tensor): it is possible to find a tetrad where some of these scalars vanish, depending on how algebraically special the $n$-spinor is. A summary of the classification for the field strength tensor is given in table \ref{4dpetrovfieldstrength}, and for the Weyl tensor in table \ref{4dpetrovweyl}. The Petrov scalars have the interesting property that it is always possible to choose a tetrad where $\phi_0$ vanishes. This turns out to not always be true for higher dimensions, as originally found by CMPP in \cite{Coley2004a}.
 
 \begin{table}[h]
 	\centering
 	\begin{tabular}{|c c l |} 
 		\hline
 		 Type & Spinor Alignment &Scalars   \\ 
 		\hline
 		Type I &  11& $\phi_0=0$  \\ 
 		Type II & \underline{11} & $\phi_0=\phi_1=0$  \\
 		\hline
 	\end{tabular}
 	\caption{Table showing the Petrov classes of a 2-form. There are two possible classes, only one of which is algebraically special. We denote spinor alignment, i.e., when two spinors are the same (up to scale), by underlining them. Note that the scalars only vanish in certain tetrads.}
 	\label{4dpetrovfieldstrength}
 \end{table}
 
 \begin{table}[h]
 	\centering
 	\begin{tabular}{|c c l |} 
 		\hline
 		Type & Spinor Alignment &Scalars  \\ 
 		\hline
 		Type I &  1111& $\psi_0=0$ \\ 
 		Type II & \underline{11}\,11 & $\psi_0=\psi_1=0$  \\
 		Type D &   \underline{11}\;\underline{11}& $\psi_0=\psi_1=\psi_3=\psi_4=0$  \\ 
 		Type III & \underline{111}\,1 & $\psi_0=\psi_1=\psi_2=0$  \\
 		 Type N & \underline{1111} & $\psi_0=\psi_1=\psi_2=\psi_3=0$  \\
 		\hline
 	\end{tabular}
 	\caption{Table showing the Petrov classes of a Weyl tensor. There are four different algebraically special classes. The spinor alignment indicates when two or more spinors are the same by underlining them, for example  \underline{11}\;\underline{11} refers to two different pairs of identical spinors. Note that the scalars only vanish in certain tetrads. For completeness, we note that, beyond the types represented in the table, there is also type O corresponding to a vanishing Weyl tensor. Henceforth, we will not consider explicitly this trivial type O case.}
 	\label{4dpetrovweyl}
 \end{table}
 
 Before proceeding, let us point out that the Weyl spinor, as a totally symmetric rank-4 spinor, can always be decomposed in terms of four rank-1 spinors as
  \begin{equation}
\Psi_{\alpha \beta\gamma\delta} = \alpha_{(\alpha}\beta_{\beta}\gamma_{\gamma}\delta_{\delta)}\,.
 \end{equation}
 This decomposition allows for an alternative viewpoint on the Petrov classification. The distinct algebraic classes are given by the alignment of the rank-1 spinors , i.e., the equivalence of the rank-1 spinors up to scale. We have represented the aligned spinors in tables \ref{4dpetrovfieldstrength} and \ref{4dpetrovweyl} by underlining them.

The reduction of the four-dimensional formalism reviewed in this section to three dimensions is discussed in \cite{Milson:2012ry}.

\section{A Newman-Penrose basis in five dimensions}
\label{spinors}

In the study of scattering amplitudes, it is important to construct a basis of vectors associated with a given particle. Physically, these vectors are the momenta of a particle, a choice of gauge, and a basis of polarisation vectors. A method to construct this basis, known as the spinor-helicity method, is known in any dimension~\cite{Cheung2009,CaronHuot:2010rj,Boels:2012ie}. The method builds on foundational work on amplitudes in four dimensions~\cite{XuZhangChang,GunionKunszt,KleissStirling,DeCausmaecker:1981by,Berends:1981rb}. 

In four dimensionals, the spinor-helicity construction is reminiscent of the Newman-Penrose tetrad, suggesting that the spinor-helicity method can be adapted to craft a higher-dimensional Newman-Penrose basis. We will see below that this turns out to be the case, focusing on five dimensions for concreteness. Apart from some comments on six dimensions in section~\ref{highdim}, we leave higher dimensions for future work.

We begin with five-dimensional flat space. We will generalise to curved space in section~\ref{Section::spinors in curved space}. 
		
\subsection{Spinors in five dimensions}
\label{sec:5dspinors}

Our five-dimensional setup is based on the six-dimensional conventions of~\cite{Cheung2009}, taking into account simplifications which occur in odd dimensions~\cite{Boels:2012ie}. Even dimensions always have the property that one can choose a chiral basis of $\gamma$ matrices, leading to the Clifford algebra\footnote{In even dimensions, there is always a matrix $\gamma_*$ with the property that $\{\gamma^\mu, \gamma_*\} = 0$. In four dimensions, this $\gamma_*$ is usually denoted $\gamma_5$. With the help of $\gamma_*$, one can define projectors $P_\pm = (1 \pm \gamma_*) / 2$. Spinors which are eigenstates of these projectors are called chiral. The Clifford algebra $\sigma^\mu \tilde \sigma^\nu + \sigma^\nu \tilde \sigma^\mu = -2 \eta^{\mu\nu}$ can be obtained from the usual Dirac gamma algebra by defining $\sigma^\mu = P_+ \gamma^\mu P_-$ and $\tilde \sigma^\mu = P_- \gamma^\mu P_+$.}.
But in odd dimensions no such chiral choice exists. We therefore work with a basis of five $\gamma$ matrices. One can always raise and lower indices of $\gamma$ matrices; see e.g.~\cite{VanProeyen:1999ni} for a useful review. In five dimensions, we may also exploit the accidental isomorphism between $\mathfrak{so}(5)$ and $\mathfrak{sp}(2)$ to choose our $\gamma$ basis so that the matrices with lower indices are antisymmetric. Since it is convenient to understand the dimensional reduction to four dimensions, we found it useful to pick an explicit basis given by
\begin{equation}\label{gammabasis1}
	\gamma^{\hat{\mu}}{}_{AB}=\begin{pmatrix}
	0&\sigma^{\hat{\mu}}{}^\alpha{}_{\dot{\beta}}\\-\tilde{\sigma}^{\hat{\mu}}{}_{\dot{\alpha}}{}^\beta&0
	\end{pmatrix},\hspace{10pt}{\hat{\mu}}=0,1,2,3,
\end{equation}
where the matrices $\sigma$ and $\tilde \sigma$ are nothing but the four-dimensional Clifford bases given in equations~\eqref{eq:sigma4def} and~\eqref{eq:tsigma4def} with their spinor indices appropriately raised or lowered. The final component of the basis, $\gamma^4{}_{AB}$, is chosen to be
\begin{equation}\label{gammabasis2}
	\gamma^4{}_{AB}=-i\begin{pmatrix}
	\epsilon^{\alpha\beta}&0\\0&\epsilon_{\dot{\alpha}\dot{\beta}}
	\end{pmatrix}.
\end{equation}

With this choice of basis, we may build on our understanding of the four-dimensional NP tetrad to lay the foundations of a five-dimensional formalism. To do so, we pick null vectors $k$ and $n$ satisfying $k \cdot n \neq 0$, and choose a coordinate system in which $k^\mu$ and $n^\mu$ take the form
\begin{equation}
	k^\mu=(k^0,k^1,k^2,k^3,0),\hspace{10pt} 	n^\mu=(n^0,n^1,n^2,n^3,0).
\end{equation}
Without loss of generality, we may choose $k \cdot n = -1$. We emphasise that this choice is not necessary. It is merely a choice that allows us to explicitly incorporate familiar four-dimensional expressions. The final formulae, which are summarised in section~\ref{subsec:summary} for convenience, do not depend on this choice of components. In the following, $k$ and $n$ will be elements of a five-dimensional basis of vectors, which we will complete shortly in section~\ref{Section::Polarisationvectors}.

Our first task, however, is to construct a basis of the space of spinors in five dimensions. As in the four-dimensional case described in section~\ref{sec:4dNP}, we will find this basis by solving the massless Dirac equations for the null vectors $k$ and $n$.

Let us take $k^\mu$ as an example. We must find the null space of the matrix
\begin{equation}
	k\cdot\gamma_{AB}= \begin{pmatrix}
	0&k\cdot\sigma{}^{{\alpha}}{}_{\dot{\beta}}\\-k\cdot\tilde{\sigma}\,{}_{\dot{\alpha}}{}^\beta&0
	\end{pmatrix}.
\end{equation}
Since $k\cdot\sigma$ and $k\cdot\tilde{\sigma}$ have rank one, the matrix $k\cdot\gamma$ has rank two and the null space is two-dimensional. We conclude that the null space of $k\cdot\gamma_{AB}$ is spanned by the spinors
\begin{equation}
	\sk^A{}_1=\begin{pmatrix}
0 \\\tilde {o}^{\dot{\alpha}}
	\end{pmatrix},\hspace{20pt}
		\sk^A{}_2=\begin{pmatrix}
		o_\alpha \\0
		\end{pmatrix},
\end{equation}
which are evidently linearly independent and lie in the null space by virtue of the definitions, equations~\eqref{dirac} and~\eqref{tdirac}, of $o$ and $\tilde o$. It is very convenient to package these spinors up using a Roman two-dimensional index $a$:
\begin{equation}
	\sk^A{}_a=\begin{pmatrix}
	0&o_\alpha \\\tilde {o}^{\dot{\alpha}}&0
	\end{pmatrix}.
\end{equation}
We will see below that the spinors $\sk^A{}_1$ and $\sk^A{}_2$ transform into one another under the action of a particular group.

To get a feel for $\sk^A{}_a$, it is helpful to understand its relationship with the vector $k^\mu$. The simplest way we can construct a spacetime vector is to hook up the indices as $\,\sk_a \sdot	\gamma^\mu \sdot \sk{}^a\,$, where we use  $\sdot$ to denote the contraction of $SO(4,1)$ spinor indices, and have defined $\sk^a = \epsilon^{ab} \sk_b$. This turns out to be correct:
for the first four components ${\hat{\mu}}=0,1,2,3$, we find	\begin{equation}
	\begin{split}
\sk_a \sdot	\gamma^{\hat{\mu}} \sdot \sk{}^a =&  \Tr \left[
\begin{pmatrix}
0&\tilde {o}^{\dot{\alpha}}\\o_\alpha &0
\end{pmatrix}\begin{pmatrix}
0&\sigma^{\hat{\mu}}{}^\alpha{}_{\dot{\beta}}
\\
-\tilde{\sigma}^{\hat{\mu}}{}_{\dot{\alpha}}{}^\beta&0
\end{pmatrix} 
\begin{pmatrix}
o_\beta&0
\\
0&- \tilde {o}^{\dot{\beta}}
\end{pmatrix}
\right]
\\ = 
&\; \sigma^{\hat{\mu}}{}_{\alpha\dot{\beta}}\,o^\alpha \,\tilde {o}^{\dot{\beta}}+ \tilde{\sigma}^{\hat{\mu}}{}_{\dot{\alpha}{\beta}}\,\tilde {o}^{\dot{\alpha}}\,
o^\beta 
\\
=&\; 2\sqrt{2}\, k^{\hat{\mu}},
	\end{split}
	\end{equation}
	while for the final component we find
\begin{equation}
\begin{split}
\sk_a \sdot	\gamma^{4} \sdot \sk{}^a =&\;-i 
\, \Tr \left[
\begin{pmatrix}
0&\tilde {o}^{\dot{\alpha}}\\o_\alpha &0
\end{pmatrix}\begin{pmatrix}
\epsilon^{\alpha\beta}&0\\0&\epsilon_{\dot{\alpha}\dot{\beta}}
\end{pmatrix}
\begin{pmatrix}
o_\beta&0
\\
0&- \tilde {o}^{\dot{\beta}}
\end{pmatrix}
\right]
 =  0.
\end{split}
\end{equation}	
Thus, using only the four-dimensional definitions, we have recovered $k^\mu=(k^{\hat{\mu}},0)$. The complete formula is therefore:
\begin{equation}
	k^\mu=\frac{1}{2\sqrt{2}}\,\sk_a \sdot \gamma^\mu \sdot \sk{}^a.
\end{equation}

It is worth commenting further on this formula. The spinors $\sk_a$ for $a = 1,2$ are a basis of solutions of the equation $k \cdot \gamma_{AB} \, \sk^B{}_a = 0$. We may, of course, perform a complex linear change of basis in this space of solutions. The normalisation condition $k^\mu=\frac{1}{2\sqrt{2}} \, \sk_a \sdot \gamma^\mu \sdot \sk{}^a$ restricts this change of basis to be an element of $SL(2, \mathbb{C})$, so we can think of the null space as a two-dimensional representation of $SL(2, \mathbb{C})$. In fact, we will see below in section~\ref{realityconditions} that if we choose a real vector $k^\mu$, and impose both our normalisation condition and a reality condition on the spinors $\sk_a$, we must further restrict this group to $SU(2)$. The physical role of this group is simply the three-dimensional rotations on the spacetime dimensions orthogonal to both $k$ and $n$.

Now we construct the other half of the spinor basis $\sn^A{}_a$. In view of the normalisation condition $k \cdot n = -1$ satisfied by the vectors, we can choose the spinors $\sk^A{}_a$ and $\sn^A{}_a$ to satisfy $\sk_a\sdot \sn_b \equiv \sk^A{}_a \Omega_{AB} \sn^B{}_b = \epsilon_{ab}$, where the raising/lowering matrix $\Omega_{AB}$ is, explicitly,
\begin{equation}
	\Omega_{AB}=\begin{pmatrix}
	\epsilon^{\alpha
		\beta}&0\\0&-\epsilon_{\dot{\alpha}\dot{\beta}}
	\end{pmatrix}.
\end{equation}
Incidentally, for notational simplicity we define
\begin{equation}
	\sk_A{}_a=\Omega_{AB}\,\sk^B{}_a,\hspace{10pt} 	\sn_A{}_a=\Omega_{AB}\,\sn^B{}_a.
\end{equation}
Following the recipe described above we find a basis of spinors in the null space of $n\cdot \sigma_{AB}$. However, a naive application of the method leads to a basis which does not satisfy our normalisation condition $\sk_a\sdot \sn_b = \epsilon_{ab}$. To correct this, we simply perform a change of basis, finding
\begin{equation}
	\sn^A{}_a =\begin{pmatrix}
	\imath_\alpha&0\\ 0&-\tilde {\imath}{}^{\dot{\alpha}}
	\end{pmatrix}.
\end{equation}
The spacetime vector $n^\mu$ can be reconstructed from the spinors as before:
\begin{equation}
	n^\mu=\frac1{2\sqrt{2}} \, \sn_a\sdot \gamma^\mu \sdot \sn{}^a.
\end{equation}The other two contractions are $\sk_a\sdot \sk_b = \sn_a\sdot \sn_b = 0$, which follows from the antisymmetry of $\Omega_{AB}$.

\subsection{Polarisation vectors}\label{Section::Polarisationvectors}

The spinors $\sk^A{}_a$ and $\sn^A{}_a$ are a complete basis of spinors. As in the four-dimensional case, we can use the spinorial basis to construct vectors which, accompanied by $k^\mu$ and $n^\mu$, form a complete basis of vectors in five dimensions -- a pentad.
Recall that the vectors $k^\mu$ and $n^\mu$ are given by
\begin{equation}\label{kndefns}
	k^\mu=\frac1{2\sqrt{2}}\, \sk_a \sdot \gamma^\mu \sdot \sk^a,\hspace{10pt}	n^\mu=\frac1{2\sqrt{2}}\, \sn_a \sdot \gamma^\mu \sdot \sn^a.
\end{equation}
We define the remaining independent contraction to be
\begin{equation}\label{polarisationdefn}
\varepsilon^\mu{}_{ab}\equiv \sk_a \sdot \gamma^\mu \sdot \sn_b=- \sn_b \sdot \gamma^\mu \sdot \sk_a
\end{equation}where it can be shown that $\varepsilon^{\mu}{}_{ab}=\varepsilon^{\mu}{}_{ba}$ by use of gamma matrix algebra. Thus, the three independent vectors associated with $\varepsilon^{\mu}{}_{ab}$ complete the pentad.

We can show this explicitly with our previous choice of components. Firstly, we will consider ${\hat{\mu}}=0,1,2,3$. For these values of ${\hat{\mu}}$, $\varepsilon^{\hat{\mu}}{}_{ab}$ is given by:
\begin{equation}
\begin{split}
\varepsilon^{\hat{\mu}}{}_{ab} = &\begin{pmatrix}
0& \tilde {o}{}^{\dot{\alpha}}\\o_\alpha&0
\end{pmatrix}\begin{pmatrix}
0&\sigma^{\hat{\mu}}{}^\alpha{}_{\dot{\beta}}\\-\tilde{\sigma}^{\hat{\mu}}{}_{\dot{\alpha}}{}^\beta&0
\end{pmatrix}\begin{pmatrix}
\imath_\beta&0\\ 0&-\tilde {\imath}{}^{\dot{\beta}}
\end{pmatrix}
\\=& \begin{pmatrix}
\tilde{\sigma}^{\hat{\mu}}{}_{\dot{\alpha}\beta}\,\tilde {o}^{\dot{\alpha}}\, \imath^\beta&0\\0&\sigma^{\hat{\mu}}{}_{\alpha\dot{\beta}}\,o^\alpha\,\tilde {\imath}^{\dot{\beta}}
\end{pmatrix}
\\=&
\sqrt{2}\, \begin{pmatrix}
{m}^{\hat{\mu}}&0\\0&\widetilde{m}^{\hat{\mu}}
\end{pmatrix}.
\end{split}
\end{equation}
Thus we can see that as long as $\hat{\mu}=0,1,2,3$, the diagonal components of $\varepsilon^{\hat{\mu}}{}_{ab}$ are precisely the vectors $m^{\hat{\mu}}$ and $\widetilde{m}^{\hat{\mu}}$ which appeared in the Newman-Penrose tetrad in four dimensions. The final value of $\mu$, $\mu=4$, is given by
\begin{equation}
\begin{split}
\varepsilon^4{}_{ab}=&\;\sk_a \sdot \gamma^4 \sdot \sn_b
\\
=&\;
-i\begin{pmatrix}
0& \tilde {o}{}^{\dot{\alpha}}\\o_\alpha&0
\end{pmatrix}\begin{pmatrix}
\epsilon^{\alpha\beta}&0\\0&\epsilon_{\dot{\alpha}\dot{\beta}}
\end{pmatrix}\begin{pmatrix}
\imath_\beta&0\\ 0&-\tilde {\imath}{}^{\dot{\beta}}
\end{pmatrix}
\\=&\;
\begin{pmatrix}
0&i\\i&0
\end{pmatrix}.
\end{split}
\end{equation}
We therefore find
\begin{equation}
\begin{split}
 \varepsilon^\mu{}_{11}= &\;\sqrt{2}\, \left({m}^{\hat{\mu}} ,0\right)
 \\\varepsilon^\mu{}_{22}= &\;\sqrt{2}\, \left(\widetilde{m}^{\hat{\mu}},0\right)
 \\\varepsilon^\mu{}_{12}=\varepsilon^\mu{}_{21}= &\;  \left(0,0,0,0,i\right).
\end{split}
\end{equation}  Finally, we can establish the useful property
\begin{equation}
\varepsilon^\mu{}_{ab} \,\varepsilon_{\mu\, cd}=\epsilon_{ac}\,\epsilon_{bd}+\epsilon_{ad}\,\epsilon_{bc}
\end{equation} by explicit computation. The spinorial completeness relations imply that
\begin{equation}
\begin{split}
	\eta^{\mu\nu}=&\;-k^\mu n^\nu -k^\nu n^\mu + \frac12 \, \epsilon^{ac}\, \epsilon^{cd}\, \varepsilon^\mu{}_{ab}\, \varepsilon^\nu{}_{cd}.
\end{split}
\end{equation}
These properties are characteristic of polarisation vectors, which in part accounts for the utility of this formalism in scattering amplitudes.

\subsection{Reality conditions}\label{realityconditions}

Our $\gamma$ basis satisfies
\begin{equation}
(\gamma^\mu)^\dagger = -H \sdot\gamma^\mu \sdot H^T
\end{equation}
where the matrices $\gamma^\mu$ have lower indices and
\begin{equation}
H = \begin{pmatrix} 0 & \epsilon^{\dot \alpha \dot \beta} \\ - \epsilon_{\alpha \beta} & 0 \end{pmatrix} .
\end{equation}
For a real null vector $V$ in five-dimensional Minkowski space, we may impose a reality condition on the associated basis of spinors $\lambda^A{}_a$. Regarding $\lambda$ as a four-by-two matrix, reality of $V$ implies
\begin{equation}
V \cdot \gamma \sdot \lambda = 0 
\Rightarrow V  \cdot \gamma \sdot H^T \sdot \lambda^* = 0.
\end{equation}
Thus the spinors $H^T \sdot \lambda^*$ are linear combinations of the two basis spinors $\lambda_a$, so we may write $H^T\sdot \lambda^* = \lambda X$, where $X$ is a two-by-two matrix.

Recall from section~\ref{sec:5dspinors} that the two-dimensional space of $\lambda_a$ furnishes a representation of $SL(2, \mathbb{C})$. The reality condition $H^T \sdot\lambda^* = \lambda X$ is not covariant under the full $SL(2, \mathbb{C})$, because the left-hand side transforms under the conjugate representation of the right-hand side. Thus the group is broken to $SU(2)$, which has the well-known property that the conjugate representation is equivalent to the fundamental representation. Requiring that the reality condition is covariant under this $SU(2)$ determines $X \propto \epsilon$. Thus, in our conventions, we arrive at the reality condition in the form~\cite{Kugo:1982bn}
\begin{equation}
H^T \sdot\lambda^* = -\lambda \epsilon.
\end{equation}
Using index notation, we may write this as follows. First we define $\bar \lambda^{\dot A a} \equiv (\lambda^A{}_a)^*$; then the reality condition is
\begin{equation}
\bar \lambda^{\dot A a} H_{\dot A}{}^A = \epsilon^{ab} \lambda^A{}_b.
\label{eq:reality}
\end{equation}

Our main focus will be on real spacetimes with Minkowski signature. Therefore we will pick real vectors $k^\mu$ and $n^\mu$ and impose the reality condition, equation~\eqref{eq:reality}, on the spinors $\sk^A{}_a$ and $\sn^A{}_a$.

We must now investigate what this means for our pentad, in particular for the ``polarisations''  $\varepsilon^\mu{}_{ab}$. They are defined by $\varepsilon^\mu{}_{ab}~=~\sk_a \sdot \gamma^\mu \sdot \sn_b$; we define the conjugate of these vectors to be $\overline{\varepsilon}^\mu{}^{ab} \equiv (\varepsilon^\mu{}_{ab})^*$. Using the reality condition we find
\begin{equation}
\begin{split}
\bar{\varepsilon}^\mu{}^{ab}&= (\sk_a \sdot \gamma^\mu \sdot \sn_b)^* \\
&= (\sk_a)^*\sdot (\gamma^\mu)^* \sdot(\sn_b)^* \\
&= \bar \sk^a \sdot( H \sdot\gamma^\mu \sdot H^T) \sdot \bar \sn^b \\
&= (\epsilon^{ac}\, \sk_c) \sdot \gamma^\mu \sdot (\epsilon^{bd} \,\sn_d) \\
&= \epsilon^{ac}\, \epsilon^{bd}\, \varepsilon^\mu{}_{cd} \\
&= \varepsilon^\mu{}^{ab}.
\end{split}
\end{equation}
In short, $\varepsilon^\mu{}^{ab}= \left(\varepsilon^\mu{}_{ab}\right)^*$. So $\varepsilon^\mu{}_{11}=(\varepsilon^\mu{}_{22})^*$, while $\varepsilon^\mu{}_{12}=-(\varepsilon^\mu{}_{12})^*$. This is exactly as we found in section \ref{Section::Polarisationvectors}: $\varepsilon^\mu{}_{11}$ and $\varepsilon^\mu{}_{22}$ relate to $m^\mu{}$ and $\widetilde{m}^\mu{}$ respectively while $\varepsilon^\mu{}_{12}$ is given by $i e^\mu{}_4$, which is indeed imaginary.

	\subsection{Lorentz transformations and the little group}\label{lorentztransforms}

To build some intuition into the objects $\sk^A{}_a$ and $\sn^A{}_a$, it is worth pausing our development to understand how these spinors transform under symmetries, especially (local) Lorentz transformations. Recall that the index $A$ takes values from 1 to 4, spanning the four dimensions of the spinorial representation of $SO(4,1)$, while the index $a$ takes values 1 and 2 and spans the two-dimensional solutions space of, for example, the equation $k_\mu \gamma^\mu_{AB} \sk^B_a = 0$. We will see that the $SU(2)$ acting on the two-dimensional solution space is the subgroup of Lorentz transformations which preserve the vector $k^\mu$. This subgroup is the little group of the null vector $k^\mu$.
	
	\subsubsection{Boosts and spins}

We have defined the spinors $\sk^A{}_a$ and $\sn^A{}_a$ to be solutions of the Dirac equations $k\cdot \gamma_{AB} \sk^B_a = 0 = n\cdot \gamma_{AB} \sn^B_a$, subject to the normalisation condition $\sk_a\cdot \sn_b = \epsilon_{ab}$, and obeying a reality condition for real spacetimes. Obviously the rescaling
\begin{equation}
\sk^A{}_a\rightarrow b \, \sk^A{}_a, \quad\quad \sn^A{}_a\rightarrow \frac{1}{b} \,\sn^A{}_a
\label{eq:Lboost}
\end{equation}
will preserve the definitions, provided that the factor $b$ is real for real spacetimes. We may therefore investigate how this rescaling acts on the pentad we have constructed from the spinors, equations~\eqref{kndefns} and~\eqref{polarisationdefn}. It is easy to see that the action is
\begin{equation}
\label{eq:boostgr}
	k^\mu\to b^2 \, k^\mu,
	\quad\quad n^\mu\to \frac{1}{b^2}\,n^\mu,
	\quad\quad \varepsilon^\mu{}_{ab}\to \varepsilon^\mu{}_{ab}.
\end{equation}
This simple transformation is nothing but a Lorentz boost in the two-dimensional space spanned by $k^\mu$ and $n^\mu$, leaving the remaining three dimensions invariant.

We may also consider a more non-trivial change of basis of the solution space of the Dirac equations:
\begin{equation}
\sk^A{}_a \rightarrow \sk'^A{}_a = M_a{}^b\,  \sk^A{}_b, \quad\quad \sn^A{}_a \rightarrow \sn'^A{}_a= N_a{}^b \sn^A{}_b.
\end{equation}
This change of basis automatically preserves the conditions that $\sk_a \sdot \sk_b = 0$ and $\sn_a \sdot \sn_b = 0$. We have already seen that $M$ and $N$ are elements of $SL(2, \mathbb{C})$. The normalisation condition is that
\begin{equation}
M_{a}{}^{c}\, N_{b}{}^{d} \, \epsilon\,_{cd}=\epsilon_{ab},
\label{eq:defOfN}
\end{equation} 
which implies that $N = M$.

We may now investigate the action of this group of transformations on our spacetime pentad. A straightforward calculation shows that the transformation is
\begin{equation}
	k^\mu\to k^\mu, \quad\quad n^\mu\to n^\mu, \quad\quad \varepsilon^\mu{}_{ab}\to M_a{}^c \, M_b{}^d\,  \varepsilon^\mu{}_{cd}.
\end{equation}
This is a Lorentz transformation preserving $k$ and $n$.

In the real case, we have already seen that the transformation $M$ is an element of $SU(2)$. This makes sense: in the real case, the subgroup of the Lorentz group which preserves $k^\mu$ and $n^\mu$ is evidently $SO(3)$. We can see this more concretely by introducing a vectorial basis of the three-dimensional representation of $SU(2)$, which is also the fundamental representation of $SO(3)$. The symmetric Pauli matrices\footnote{The usual Pauli matrices are $-2i\epsilon_{ac}\,\varsigma^{cb}{}_{i}$.} $\varsigma^{ab}{}_{i}$, $i=1,2,3$ provide a convenient mapping from the $\underline{2} \otimes \underline{2}$ tensor product of $SU(2)$ representations to the $\underline{3}$. In view of the reality condition, we find it convenient to take
\begin{equation}
\varsigma_1 = \frac 12 \begin{pmatrix} i & 0 \\ 0 & -i \end{pmatrix}, \quad\quad
\varsigma_2 = \frac 12 \begin{pmatrix} 1 & 0 \\ 0 & 1 \end{pmatrix}, \quad\quad
\varsigma_3 = \frac 12 \begin{pmatrix} 0 & -i \\ -i & 0 \end{pmatrix}.
\end{equation}
Notice, for example, that this choice of basis has the property that $(\varsigma^{11}{}_i)^* = \varsigma^{22}{}_i$, consistent with our reality condition.

We may then define
\begin{equation}
\label{eq:5dpolvec}
\varepsilon^{\mu}{}_{ i} = \varepsilon^\mu{}_{ab} \, \varsigma^{ab}{}_{i},
\end{equation}
and 
\begin{equation}
	m_i = M_{ab} \, \varsigma^{ab}{}_{i}.
\end{equation}
The antisymmetric degree of freedom in $M$ is defined to be $M_{\mathrm{tr}}=\epsilon^{ab} M_{ab}$. In this language, the condition that $M$ has unit determinant becomes $\frac{1}{4}M_{\mathrm{tr}}{}^2 + \underline{m} \cdot \underline{m}=1$, and the polarisation vector transformation is
\begin{equation}
\underline{\varepsilon}^\mu\to \left( -\underline{m} \cdot \underline{m} +\frac14 M_\mathrm{tr}{}^2\right)\underline{\varepsilon}^\mu +2 \left(\underline{m} \cdot\underline{\varepsilon}^\mu \right) \underline{m}+M_\mathrm{tr} \left(\underline{m}\times\underline{\varepsilon}^\mu\right).
\end{equation}
We can compare this with the standard formula for a rotation by angle $\theta$ around an axis $\underline{n}$ in three-dimensional Euclidean space,
\begin{equation}
\underline{x}\rightarrow \cos\theta \, \underline{x} +\left(1-\cos\theta\right)\, \left(\underline{n}\cdot \underline{x}\right)\, \underline{n} + \sin\theta \, \left(\underline{n}\times\underline{x}\right),
\end{equation}
to see that the transformation $M$ rotates the polarisation vectors by an angle $\sin\theta=M_{\mathrm{tr}}|\underline{m}|$ around the axis $\underline{m}$ in the Euclidean 3-space of the little group, leaving $k^\mu$ and $n^\mu$ invariant. 

\subsubsection{The null rotations }

The boost and spin transformations comprise four of the ten Lorentz transformations available in a five-dimensional spacetime. It is interesting to understand the remaining six. To do so, we look to the null rotations of the four-dimensional NP tetrad for inspiration, and construct the ansatz $\sk^A{}_a\rightarrow \sk^A{}_a + T_a{}^b\, \sn^A{}_b,$ $\sn^A{}_a\rightarrow \,\sn^A{}_a$. To preserve $\sk_a\cdot \sn_b$, we require that the matrix $T$ is symmetric:
\begin{equation}
\begin{split}
\sk_a' \cdot \sk_b ' = &\, \left(\sk_a +T_a{}^c \,\sn_c \right)\cdot \left(\sk_b +T_b{}^d\, \sn_d \right)
\\=& \; T_a{}^c \left( \sn_c \cdot \sk_b \right) +T_b{}^d  \left(\sk_a \cdot \sn_d \right)
\\=&  \; T_{ab} - T_{ba} =0.
\end{split}	
\end{equation}
Similarly the transformation $\sk^A{}_a\rightarrow \sk^A{}_a$, $
\sn^A{}_a\rightarrow \sn^A{}_a+ S_a{}^b\, \sk^A{}_b$ is valid as long as $S$ is symmetric. The symmetric matrices $S$ and $T$ comprise three degrees of freedom each, so combined with the boost and spin, this is a complete parametrisation of the Lorentz group. The action of these transformations on our pentad is:
	\begin{itemize}
		\item Null rotation about $n$:  \,  $\sk^A{}_a\rightarrow \sk^A{}_a + T_a{}^b\, \sn^A{}_b,\hspace{10pt}\sn^A{}_a\rightarrow \sn^A{}_a$,	\begin{equation}
		 k^\mu\to k^\mu + T^{ab} \; \varepsilon^\mu{}_{ab} - \det T\,  n^\mu	
		 ,\hspace{10pt}n^\mu\to n^\mu
				,\hspace{10pt} \varepsilon^\mu{}_{ab}\to \varepsilon^\mu{}_{ab} +T_{ab} \,n^\mu.
		\end{equation}
		
		\item Null rotation about $k$: \,	$	\sk^A{}_a\rightarrow \sk^A{}_a ,\hspace{10pt}
		\sn^A{}_a\rightarrow \sn^A{}_a+ S_a{}^b\, \sk^A{}_b$,	\begin{equation}
			k^\mu\to k^\mu
			,\hspace{10pt} n^\mu\to n^\mu + S^{ab}\, \varepsilon^\mu{}_{ab} - \det S\, k^\mu
			,\hspace{10pt} \varepsilon^\mu{}_{ab}\to \varepsilon^\mu{}_{ab} + S_{ab}\, k^\mu.
		\end{equation}

\end{itemize}

\subsection{Summary}
\label{subsec:summary}

We can now summarise the key results. The pentad is constructed from the null orthogonal vectors $k^\mu$ and $n^\mu$, satisfying
\begin{equation}
	k^2=n^2=0, \quad \quad k_\mu \,n^\mu = -1,
\end{equation}
and from the three independent spacetime vectors contained in the symmetric polarisation vector $\varepsilon^\mu{}_{ab}$,
satisfying
\begin{equation}
	k\cdot\varepsilon_{ab}=n\cdot\varepsilon_{ab}=0,\quad\quad 	\varepsilon_{\mu ab}\,\varepsilon^\mu{}_{ cd}=\epsilon_{ac}\,\epsilon_{bd}+\epsilon_{ad}\,\epsilon_{bc}.
\end{equation}
This pentad spans the spacetime as 
\begin{equation}
\begin{split}
\eta^{\mu\nu}=&\;-k^\mu n^\nu -k^\nu n^\mu + \frac12 \, \epsilon^{ac}\, \epsilon^{cd}\, \varepsilon^\mu{}_{ab}\, \varepsilon^\nu{}_{cd}.
\end{split}
\end{equation}

We choose spinors $\sk^A{}_a$, $\sn^A{}_a$, where $A=1,...,4$ is a spacetime spinor index and $a=1,2$ is a little group spinor index, to satisfy
\begin{equation}
 \sk_a \sdot \sk_{b}=\sn_a \sdot \sn_{b}=0,\quad\quad \sk_a \sdot \sn_{b}= \epsilon_{ab},
\end{equation}
where ``$x\sdot y $'' indicates a contraction on the spacetime spinor index, i.e., $x_A\, y^A$. The pentad can be defined in terms of the spinors:
\begin{equation}\label{NormalisationRelations}
\begin{split}
	k^\mu=&\;\frac{1}{2\sqrt{2}} \; \sk_a \sdot \gamma^\mu \sdot \sk{}^a,\quad\quad
	n^\mu=\frac1{2\sqrt{2}} \; \sn_a\sdot \gamma^\mu \sdot \sn{}^a,\quad\quad
\varepsilon^\mu{}_{ab}= \sk_a \sdot \gamma^\mu \sdot \sn_b,\quad
\end{split}
\end{equation}
in order to automatically satisfy the properties given above. 
To restrict to real Minkowski space, the spinors must satisfy reality conditions. In particular, any real objects which transform under the little group indices must obey
\begin{equation}
	\left(X_{a_1 ... a_n}{}^{b_1 .... b_m} \right){}^* = X^{a_1 ... a_n}{}_{b_1 .... b_m}.
\end{equation}
Finally, we note that the ten transformations of the standard five-dimensional Lorentz group can be parametrised as a boost $b$, three spins $M_{ab}$ where $\det M=1$, and two three-dimensional null transformations $T_{ab}$ and $S_{ab}$ which are both symmetric:
	\begin{itemize}
	\item Boost:\,  $
	\sk_A{}^a\rightarrow b \, \sk_A{}^a  ,\hspace{10pt}
	\sn_A{}^a\rightarrow \frac{1}{b} \,\sn_A{}^a$
	
	\item Spin:\,   $	\sk_A{}^a\rightarrow M^a{}_b\, \sk_A{}^b  ,\hspace{10pt}		\sn_A{}^a\rightarrow M{}^a{}_b \sn_A{}^b$

	\item Null rotation about $n$:  \,  $\sk^A{}_a\rightarrow \sk^A{}_a + T_a{}^b\, \sn^A{}_b,\hspace{10pt}\sn^A{}_a\rightarrow \sn^A{}_a$
	
	\item Null rotation about $k$: \,	$	\sk^A{}_a\rightarrow \sk^A{}_a ,\hspace{10pt}
	\sn^A{}_a\rightarrow \sn^A{}_a+ S_a{}^b\, \sk^A{}_b$ .
	\end{itemize}

\section{The field strength tensor}
\label{maxwell}

Although our main goal is to apply the results of section \ref{spinors} to gravity, it is helpful to apply them to the simpler field strength tensor  $F_{\mu\nu}$ first.

	\subsection{Set up and classifications}\label{setupandcoarseclassfieldstrength}
To begin, we contract $F_{\mu\nu}$ with the rotation generator
\begin{equation}\label{RotationMatrix}
\sigma^{\mu\nu}{}_{AB}=\frac{1}{2}\left(\gamma^{\mu}{}_{AC}\,\gamma^{\nu}\,{}^C{}_{B}-\gamma^{\nu}{}_{AC}\,\gamma^{\mu}\,{}^C{}_{B}\right)
\end{equation}  to find a symmetric bi-spinor,
 \begin{equation}
 \Phi_{AB}=F_{\mu\nu}\,\sigma^{\mu\nu}{}_{AB}.
 \end{equation}
 This is analogous to the four-dimensional Newman-Penrose formalism, as described in section \ref{recap}. Now, however, upon contraction with our basis spinors, we do not obtain scalars but little group bi-spinors:
 \begin{equation}\label{phidefns}
 \Phi^{(0)}_{ab}=\Phi_{AB}\,\sk^A{}_a\,\sk^B{}_b,\hspace{10pt}	\Phi^{(1)}_{ab}=\Phi_{AB}\,\sk^A{}_a\,\sn^B{}_b,\hspace{10pt}	\Phi^{(2)}_{ab}=\Phi_{AB}\,\sn^A{}_a\,\sn^B{}_b,
 \end{equation}
 where the bracketed numbers label the little group bi-spinors according to the number of $\sn^A{}_a$ spinors they are contracted with. To begin with, we will consider complex-valued $F_{\mu\nu}$, and restrict to the real case later on.
 
In four dimensions, the Petrov classification based on the scalars defined in \eqref{4D2formdecomposition} had two classes, type I and type II, the latter of which was considered algebraically special. Type II was defined by the existence of a tetrad where both of the four-dimensional Petrov scalars $\phi_0$ and $\phi_1$ vanished; see table \ref{4dpetrovfieldstrength}. Since the scalars from equation \eqref{4D2formdecomposition} and the spinors from \eqref{phidefns} are clearly analogous, this motivates a Petrov-like classification for five dimensions, which is shown in table \ref{5dpetrovfieldstrength}. The guaranteed existence of a tetrad where $\phi_0$ vanishes is a special feature of four dimensions, and so we also require an additional ``general'' class for 2-forms in five dimensions. As we will show in section \ref{cmpp field strength}, this is exactly the original CMPP classification for the 2-form. 
 \begin{table}[h]
 	\centering
 	\begin{tabular}{|l l |} 
 		\hline
 		Type \hspace{30pt} &  Little group spinor characteristic
 		\\ 
 		\hline
 		Type G &   $\Phi^{(i)}\neq0\; \forall \; i$ 
 		\\ 
 		Type I &   $\Phi^{(0)}=0$  
 		\\ 
 		Type II &  $\Phi^{(0)}=\Phi^{(1)}=0$  
 		\\
 		\hline
 	\end{tabular}
 	\caption{Table showing a proposed Petrov-like classification for a 2-form. There are now three possible classes, two of which are analogous to four dimensions and one of which, Type G, is new to higher dimensions. }
 	\label{5dpetrovfieldstrength}
 \end{table}

The bi-spinors defined in \eqref{phidefns} are reducible, and therefore we will refer to this classification as a ``coarse'' classification. A more fine-grained classification is available if we break the bi-spinors down into their irreducible representations, namely the symmetric bi-spinor and the scalar. To do this, we will use the notation that $\phi^{(i)}$ refers to the symmetrisation of $\Phi^{(i)}$, such that $\phi^{(i)}_{ab}=\Phi^{(i)}_{(ab)}$. Since $\Phi_{AB}=\Phi_{BA}$, we can see that $\Phi^{(0)}$ and $\Phi^{(2)}$ are already symmetric, so $\phi^{(0)}=\Phi^{(0)}$ and $\phi^{(2)}=\Phi^{(2)}$. The bi-spinor $\Phi^{(1)}$ is not symmetric in general, but it is always possible to write a two-component bi-spinor as the sum of a symmetric bi-spinor and a trace term proportional to the Levi-Civita tensor\footnote{Since a two-dimensional index has only two possible values,
 \begin{equation*}
\epsilon_{a[b}\epsilon_{cd]} = 0 = \epsilon_{ab}\,\epsilon_{cd} + \epsilon_{ac}\,\epsilon_{db} + \epsilon_{ad} \,\epsilon_{bc}.
 \end{equation*} 
 Contracting this with an arbitrary bi-spinor $s^{cd}$, we obtain
 \begin{equation*}
s_{ab}-s_{ba}=\epsilon_{ab}\, s_c{}^c.
 \end{equation*}}. We will refer to this trace as $\Phi^{(1)}{}_a{}^a = \Phi^{(1)}_{\mathrm{tr}}$ such that:
 \begin{equation}
 	\Phi^{(1)}_{ab}=\phi^{(1)}_{ab} + \frac{1}{2}\, \Phi^{(1)}_{\mathrm{tr}}\, \epsilon_{ab}.
 \end{equation}
 This is simply the statement that a \textbf{4} decomposes as $\textbf{4}=\textbf{3}+\textbf{1}$ where the symmetric bi-spinor $\textbf{3}$ and the scalar \textbf{1} are both irreducible representations.
The 10 degrees of freedom in the five-dimensional field strength tensor have therefore been split up into 3 symmetric bi-spinors and a single scalar. We can write this as in table \ref{fieldstrengthlaidoutnicely},
where the terms have been organised by the dimension of their irreducible representation along the horizontal axis and by the bracketed number in the vertical direction. This fine-grained classification is sensitive to the vanishing of the columns as well as the rows. For example, a 2-form with vanishing $\phi^{(1)}_{ab}$ or $\phi^{(1)}_{\mathrm{tr}}$ is considered more special than one where both are non-zero. We will give some examples in section \ref{Section::2formExamples}.

\begin{table}[h!]
	\centering
	\setlength{\tabcolsep}{10pt}  \renewcommand\arraystretch{1.8}\begin{tabular}{|c||c||cc|}
		\hline	Reducible representation& &\textbf{3}&\textbf{1}\\\hline
		\hline$\Phi^{(0)}_{ab}$&&$\phi^{(0)}_{ab}$&
		\\$\Phi^{(1)}_{ab}$&$\Rightarrow$&$\phi^{(1)}_{ab} $& $\Phi^{(1)}_{\mathrm{tr}}$
		\\$\Phi^{(2)}_{ab}$&&$\phi^{(2)}_{ab}$&\\\hline
	\end{tabular} 
	\caption{The three little group spinors of the 2-form can be broken up into three symmetric bi-spinors, \textbf{3}, and a scalar \textbf{1}. This fine-grained structure is able to provide more detail on the nature of the 2-form than the coarse classification. For example, a type I solution with vanishing $\vtr$ is more special than one where both $\vtr$ and $\phi^{(1)}$ are non-zero.}
	\label{fieldstrengthlaidoutnicely}
\end{table}
 
In the real case, these objects are subject to the conditions $\phi^{(i)}_{ab}=\left(\phi^{(i)\,ab}\right)^*$. We can easily recast them into real vectors acted on by $SO(3)$ using the Pauli matrices $\varsigma^i{}_{ab}$:
 \begin{equation}
\left(\phin\right)^i=\phi^{(0)}_{ab}\; \varsigma^i{}^{ab},
 \end{equation}
 where $i=1,2,3$ is an $SO(3)$ index, and of course $\Phi^{(1)}_{\mathrm{tr}}$ remains a scalar. The little group irreps therefore change into a combination of 3-vectors and scalars as shown in table \ref{so3fieldstrengthtensorcontent1}.  Vector notation will be useful when making contact with the existing literature. 
 \begin{table}
	\centering
	\setlength{\tabcolsep}{10pt} \renewcommand\arraystretch{1.3}\begin{tabular}{|cc||c||cc|}
 \hline
  \multicolumn{2}{|c||}{Spinor notation}&&  \multicolumn{2}{|c|}{Vector notation}\\\hline\hline
$ \phi^{(0)}_{ab}$&&&$\underline{\phi_0}$&
 \\ $\phi^{(1)}_{ab}$& $\Phi^{(1)}_{\mathrm{tr}}$&$\leftrightarrow$&$\underline{\phi_1}$&$\Phi^{(1)}_{\mathrm{tr}}$
 \\$\phi^{(2)}_{ab}$&&&$\underline{\phi_2}$&
 \\\hline
 \end{tabular}
 \caption{The little group irreps can be written in terms of spinors or vectors by standard use of the Pauli matrices. }
 \label{so3fieldstrengthtensorcontent1}
 \end{table}

Finally, it is always possible to factorise a symmetric bi-spinor into two symmetrised spinors
\begin{equation}
	\phi_{ab}=\alpha_{(a}\,\beta_{b)}.
\end{equation}
It is natural to ask if there exists some subclassification where $\alpha=\beta$ as is the case in four dimensions. From the vectorial perspective it is easy to see that this will not be the case if we restrict ourselves to real Minkowski space. If we consider an arbitrary symmetric bispinor
\begin{equation}
\left(\underline{\phi} \right)^i= \phi_{ab}\; \varsigma^i{}^{ab} = \alpha_{a}\,\beta_{b}\, \varsigma^i{}^{ab},
\end{equation}
we can see that the modulus of this vector is given by 
\begin{equation}
	|\underline{\phi}|= \frac1{2} \, |\alpha_a\beta^a|,
\end{equation}
using $\varsigma^{i\,ab} \, \varsigma^{i\,cd} =(\epsilon^{ac}\epsilon^{bd}+\epsilon^{ad}\epsilon^{bc})/4$. Therefore, there is no non-vanishing real vector $\underline{\phi}$ such that $\alpha=\beta$, and the irreps that we describe in table \ref{fieldstrengthlaidoutnicely} cannot be broken down further. In contrast, in the complex case they can, leading to a Russian doll-like structure of nested classifications where each bi-spinor $\phi^{(i)}$ can itself be type I ($\alpha\neq\beta$) or type II ($\alpha=\beta$).

\subsection{Examples}\label{Section::2formExamples}

To be more concrete, we will discuss some simple examples: the plane wave, an electric field and a magnetic field. This will illuminate some details of the fine structure.

\subsubsection{A plane wave}\label{Section::Planewave}

The simplest solution is a plane wave which has a field strength tensor of the form
\begin{equation}
	F_{\mu\nu}= k_{[\mu} \varepsilon_{\nu]}{}^{ab} P_{ab} \,e^{ik\cdot x},
\end{equation}
where the symmetric $P_{ab}$ corresponds to an arbitrary choice of polarisation. It is natural to choose $k_\mu$ and $\varepsilon_\mu{}^{ab}$ to be elements of our pentad. Using the normalisations in equation \eqref{NormalisationRelations} we have
\begin{equation}
\begin{split}
	\Phi_{AB}=	&\; F_{\mu\nu}\sigma^{\mu\nu}{}_{AB}\\
	=&\; k_{[\mu} \varepsilon_{\nu]}{}^{ ab} \gamma^\mu{}_{AC}\gamma^{\nu \, C}{}_B P_{ab} \,e^{ik\cdot x}
	\\=&\; -2\sqrt2\, \sk_{(A}{}^{a} \sk_{B)}{}^{b} P_{ab} \,e^{ik\cdot x},
\end{split}
\end{equation}
and comparison with equation \eqref{phidefns} tells us that we have
\begin{equation}
\phi^{(0)}=\Phi^{(1)}=0,\hspace{10pt} \phi^{(2)}_{ab}=-2\sqrt2 P_{ab} e^{ik\cdot x}.
\end{equation}
A plane wave is therefore a type II solution under the coarse classification. Since $\phi^{(2)}$ is symmetric, it is an irreducible representation of $SU(2)$. However, it is possible that $P_{ab}=\alpha_a\alpha_b$ in the complex case, which of course describes a circularly polarised electromagnetic field.

\subsubsection{A constant electric field}\label{Section::Electric}

Our second example is a constant electric field $\underline{E}$ in the $x$ direction. Then the Maxwell spinor has the form
\begin{equation}
\Phi_{AB}= 2|\underline{E}| \sigma^{tx}{}_{AB}.
\end{equation}
We choose $k=\frac1{\sqrt{2}}(\partial_t+\partial_x)$ and $n=\frac1{\sqrt{2}}(\partial_t-\partial_x)$. Taking contractions with $\sk^A{}_a$ and $\sn^A{}_a$, we find
\begin{equation}
\phi^{(0)}=\phi^{(1)}=\phi^{(2)}=0,\quad\quad \vtr = 4 |\underline{E}|.
\end{equation}
Hence the electric field has a coarse type I classification, but the fine structure is able to pinpoint that this is more special than a general type I.

\subsubsection{A constant magnetic field}\label{Section::Magnetic}

Finally, we consider a simple magnetic field $B$ which is trivial in the $x$ direction such that $F^{\mu\nu}=B^{ij}$. 
We use the same pentad as the previous section, so $k=\frac1{\sqrt{2}}(\partial_t+\partial_x)$ and $n=\frac1{\sqrt{2}}(\partial_t-\partial_x)$. The Maxwell spinor is 
\begin{equation}
	\Phi_{AB}=B^{ij} \sigma^{ij}{}_{AB}.
\end{equation}
Taking contractions again and using the Pauli matrices $\varsigma^i{}_{ab}$ to recast $\phi^{(1)}$ as a vector, we find
\begin{equation}
	\phi^{(0)}=\phi^{(2)}=\vtr=0,\quad\quad \left(\phio\right)^i=\epsilon^{ijk} B^{jk}.
\end{equation}
Therefore, although this magnetic field and the electric field have the same coarse classification, type I, they can be differentiated by their fine structure. 

	\subsection{Relations to the literature: CMPP and de Smet}\label{cmpp field strength}
	
As we have mentioned earlier, there exist previously proposed classifications for five-dimensional spacetimes. Two of these are the classification derived by CMPP in 2004 \cite{Coley2004a,Coley2004b} and the de Smet classification proposed in 2002 \cite{deSmet2002}. We will understand both in terms of the spinorial formalism.

\subsubsection{The CMPP classification}

In their papers \cite{Coley2004a,Coley2004b}, CMPP observe that each component of the Weyl tensor in five dimensions has a boost weight when the pentad is rescaled by $\{k,\,n,\,m^{(i)}\} \rightarrow \{\rho\, k,\, \rho^{-1} \,n,\,m^{(i)}\}$ for some scalar $\rho$, where $i =2,3,4$. This boost weight is simply the power of $\rho$ by which the component of the 2-form transforms. The independent components of the 2-form have the following boost weights:
\begin{equation}
\renewcommand\arraystretch{1.2}\begin{array}{|c||c|c|c|}
\hline
\mathrm{Boost \hspace{3pt}weight}  & 1&0&-1\\ \hline
\mathrm{Component} & F_{0i} & F_{01}, F_{ij}& F_{1i}\\\hline
\end{array}\end{equation}
where the index 0 indicates a contraction with $k$, the index $1$ indicates a contraction with $n$, and a Roman index $i$ corresponds to the space-like direction $m^{(i)}$.  The CMPP $k$ and $n$ have an identical role to our own usage, so we will use the same symbols. The relevant choices of $k$ are made by demanding that $F_{0i}$ is set to zero if possible, in which case a choice of $n$ is made to also send $ F_{01}$ and $F_{ij}$ to zero if possible. Next, the boost weights are organised into a Petrov-like classification as shown in table \ref{cmpptable2form}. 
 \begin{table}[h]
 	\centering
 	\begin{tabular}{|llc |} 
 		\hline
 		Type \hspace{20pt} & Components& CMPP special?  \\ 
 		\hline
 		Type G&   $F_{0i}\neq 0$  & No\\
 		 Type I &   $F_{0i}=0$  & No\\ 
 		Type II &  $F_{0i}=F_{01}=F_{1i}=0$  &Yes\\
 		\hline
 	\end{tabular}
 	\caption{Table showing the CMPP classes of a 2-form according to which components can be found to vanish. There are three possible classes, only one of which is considered special. The pentad is chosen so that the 2-form is as special as possible.}
 	\label{cmpptable2form}
 \end{table}
 
 In order to compare our formalism with CMPP, we can simply rewrite our little group field strength tensors in terms of $F_{\mu\nu}$. Doing this, we find the simple relationships
  \begin{equation}
  \begin{split}
  F_{0i}=\frac{1}{2\sqrt{2}}
  \,\phi^{(0)}_i,\hspace{10pt} 	 	F_{01}=\frac{1}{4}\,\Phi^{(1)}_{\mathrm{tr}},\hspace{10pt}F_{ij}=\frac{1}{2} \,\epsilon_{ijk}\phi^{(1)}_k, \hspace{10pt} F_{1i}=-\frac{1}{2\sqrt{2}}\,\phi^{(2)}_i.
  \end{split}
  \end{equation}
Since each boost weight component is exactly identifiable as one of our little group irreps, the coarse classification that we introduced in section \ref{setupandcoarseclassfieldstrength} is exactly the CMPP classification as introduced in \cite{Coley2004a}. Furthermore, the bracketed number $(i)$ of a little group spinor $\Phi^{(i)}$ relates directly to its boost weight, as it would in four dimensions.

\subsubsection{The de Smet classification}

The de Smet classification \cite{deSmet2002} has a very different set up to the CMPP classification. It uses a gamma basis such as in equations \eqref{gammabasis1}, \eqref{gammabasis2} to create a symmetric field strength 2-spinor $\Phi_{AB}$, and studies its factorisation properties to create a classification. There are two cases: in de Smet notation, if the 2-form does not factorise it is a \textbf{2}, and if it does it either a ${\textbf{11}}$ or a $\textbf{\underline{11}}$, with the two factors being equal in the latter case. Let us examine this in more detail. The symmetric 2-spinor is constructed using the rotation generator as usual,
\begin{equation}
	\Phi_{AB}=F_{\mu\nu}\,\sigma^{\mu\nu}{}_{AB}.
\end{equation}
Now, the field strength polynomial $\mathcal{F}$ is constructed by contracting in an arbitrary spinor $\xi^A$, such that
\begin{equation}
	\mathcal{F}=\Phi_{AB}\,\xi^A\, \xi^B.
\end{equation}
If the original bi-spinor had the structure $\Phi_{AB}=\alpha_{(A}\,\beta_{B)}$, the polynomial will factorise. Our formalism is based on irreducible representations of $SU(2)$, namely symmetric $SU(2)$ spinors. These have the useful property that they always totally factorise. Therefore, each little group irrep will have its own de Smet structure. We can compute this by studying each of them in turn.

The field strength spinor can be expanded in terms of our little group irreps as
\begin{equation}
\Phi_{AB}=\phi^{(0)}_{ab} \,\sn_A{}^a \,\sn_B{}^b + 2\,\phi^{(1)}_{ab} \,\sn_{(A}{}^a \,\sk_{B)}{}^b + \phi^{(1)}_{\mathrm{tr}} \,\sn_{(A}{}^a\, \sk_{B)}{}_a+\phi^{(2)}_{ab} \,\sk_A{}^a \,\sk_B{}^b .
\end{equation}
As an example, let us consider a case where only $\phi^{(2)}$ is non-zero, such as the plane wave example given in section \ref{Section::Planewave}. Now, the field strength polynomial is given by
\begin{equation}\label{MaxwellPolynomialPhi2}
\begin{split}
\mathcal{F}=&\;\phi^{(2)}_{ab} \,\sk_A{}^a \,\sk_B {}^b\, \xi^A \,\xi^B
\\=& \;\alpha_{(a}
\,\beta_{b)} \,(\sk\sdot \xi)^a\,(\sk\sdot \xi)^b
\\=&\;\left[\alpha,\sk\sdot\xi\right]\left[\beta,\sk\sdot\xi\right],
\end{split}
\end{equation}
where we have defined the factorisation of $\phi^{(2)}$ to be $\phi^{(2)}_{ab}=\alpha_{(a}
\,\beta_{b)}$, and ``\;$\sdot$\;'' indicates a contraction on a spacetime spinor index, while ``$[\;\cdot\;,\;\cdot\;]$'' is a little group spinor contraction. Clearly, this is of de Smet type \textbf{11}. 

The $\phi^{(0)}$ spinor has the same structure as $\phi^{(2)}$, and therefore a 2-form for which only $\phi^{(0)}$ was non-zero would also be a \textbf{11}. However, the $\sk$ and $\sn$ structure of the $\phi^{(1)}$ component means that its field strength polynomial behaves differently. Let us consider a 2-form where only $\phi^{(1)}$ is non-zero, for example the magnetic field from section \ref{Section::Magnetic}. This would have a field strength polynomial of the form
\begin{equation}\label{MaxwellPolynomialPhi1}
\begin{split}
\mathcal{F}=&\;2\,\phi^{(1)}_{ab} \,\sn_{(A}{}^a \,\sk_{B)} {}^b\, \xi^A \,\xi^B
\\=&\; \left[\alpha,\sn\sdot\xi\right]\left[\beta,\sk\sdot\xi\right]+\left[\alpha,\sk\sdot\xi\right]\left[\beta,\sn\sdot\xi\right],
\end{split}
\end{equation}
and thus it is of de Smet type \textbf{2}. 

For a solution like the electric field in section \ref{Section::Electric}, only the  $\Phi^{(1)}_{\mathrm{tr}}$ term is non-zero. So the field strength polynomial is
\begin{equation}
\begin{split}
\mathcal{F}=&\;\Phi^{(1)}_\mathrm{tr} \, \epsilon_{ab} \,\sn_{(A}{}^a \,\sk_{B)} {}^b\, \xi^A \,\xi^B
\\=&\;  \Phi^{(1)}_{\mathrm{tr}}\left(\left[o,\sn\sdot\xi\right]\left[\imath,\sk\sdot\xi\right]- \left[o,\sk\sdot\xi\right]\left[\imath,\sn\sdot\xi\right]\right),
\end{split}
\end{equation}
where we have used the property $\epsilon_{ab}=o_a\,\imath_b-\imath_a \,o_b$ for some basis spinors $o$ and $\imath$, normalised as $o^a\, \imath_a=1$. Therefore this is also a de Smet type \textbf{2}. 

We organise the little group irreps as shown in table \ref{desmet2form}, that is, according to boost weight (along the table's vertical direction) and according to irrep dimension (along the table's horizontal direction). Then we see that each irrep corresponds to a de Smet class. Any combination of little group irreps will result in a \textbf{2}.

\begin{table}[h!]
	\centering
\renewcommand\arraystretch{1.3}\begin{tabular}{|cc||c||cc|}\hline
 \multicolumn{2}{|c||}{Little group spinors}&& \multicolumn{2}{|c|}{de Smet class}\\
\hline\hline
$\phi^{(0)}_{ab}$&&&\textbf{11}&
\\ $\phi^{(1)}_{ab}$& $\Phi^{(1)}_{\mathrm{tr}}$&$\leftrightarrow$&$\textbf{2}$&{\textbf{2}}
\\$\phi^{(2)}_{ab}$&&&\textbf{11}&
\\\hline
\end{tabular}
\caption{Each little group spinor has a predefined de Smet class.}
\label{desmet2form}
\end{table}

 As we discussed in section \ref{setupandcoarseclassfieldstrength}, in the case of complex field strength, there is a Russian doll-like secondary layer of structure, where each $\phi^{(i)}$ can itself be either type I or type II corresponding to $\alpha\neq\beta$ or $\alpha=\beta$, respectively. It is simple to read off from equation \eqref{MaxwellPolynomialPhi2} that these have distinct de Smet types \textbf{11} and \textbf{\underline{11}} respectively, in the cases of $\phi^{(0)}$ or $\phi^{(2)}$, while we can see from equation \eqref{MaxwellPolynomialPhi1} that $\phi^{(1)}$ will be \textbf{2} and \textbf{11} respectively. However,  when we restrict to real spacetimes, only the possibilities shown in table \ref{desmet2form} are possible, since the repeated case $\alpha=\beta$ is not permitted \cite{Godazgar2010}.

\section{General relativity and the Weyl tensor}\label{GR}

\subsection{Spinors in curved space}\label{Section::spinors in curved space}

So far, our analysis has been based on flat spacetime. To generalise our results to curved space, we introduce coordinate indices $\mu, \nu$ and tangent space indices $M,N$. We can then pick an arbitrary frame $e^\mu{}_M$ satisfying $g^{\mu\nu}=e^\mu{}_M \,e^\nu{}_N \,\eta^{MN}$. Both $g^{\mu\nu}$ and $\eta^{MN}$ can be expressed in terms of an NP pentad,
\begin{equation}
\begin{split}
g^{\mu\nu}=&\;-k^\mu\, n^\nu -k^\nu \,n^\mu +\epsilon^{ac}\,\epsilon^{bd}\,\varepsilon^\mu{}_{ab}\,\varepsilon^\nu{}_{cd}
\\=&\;e^\mu{}_M \,e^\nu{}_N\left( -k^M\, n^N -k^N \,n^M +\epsilon^{ac}\,\epsilon^{bd}\,\varepsilon^M{}_{ab}\,\varepsilon^N{}_{cd}\right)
\end{split}
\end{equation}
so we can read off that the curved pentad $\{k^\mu,\, n^\mu, \, \varepsilon^\mu{}_{ab}\}$ is obtained from our flat pentad $\{k^M,\, n^M, \, \varepsilon^M{}_{ab}\}$ by contraction with $e^\mu{}_M$. Similarly, the gamma basis becomes 
\begin{equation}
\gamma^\mu{}_{AB}=e^\mu{}_M \, \gamma^M{}_{AB},
\end{equation}
such that the Clifford algebra is still satisfied, exactly as for the Newman-Penrose construction in four dimensions. Notice that the index $\mu$ of previous sections should now be seen as the index $M$, and $\mu$ is henceforth a curved spacetime index.

The results we derived in section \ref{spinors} still apply for the tangent space at  each spacetime point. Thus it is possible to choose spinors of the form 
\begin{equation}
\sk^A{}_a = \begin{pmatrix}
0 &o_\alpha \\ \overline{o}^{\dot{\alpha}}&0\end{pmatrix}, \hspace{10pt} \sn^A{}_a = \begin{pmatrix}
\imath_\alpha &0\\ 0& -\overline{\imath}^{\dot{\alpha}}\end{pmatrix}
\end{equation}
where $o$ and $\imath$ are now curved space spinors of $SU(2)\times SU(2)$. Using the curved space gamma basis, we can construct the same relationships between the spinors and the pentad,
\begin{equation}
k^\mu = \frac1{2\sqrt{2}} \,\sk_a \sdot \gamma^\mu \sdot \sk^a, \hspace{10pt} n^\mu = \frac1{2\sqrt{2}} \,\sn_a \sdot \gamma^\mu \sdot \sn^a, \hspace{10pt} \varepsilon^\mu{}_{ab} = \sk_a \sdot \gamma^\mu \sdot \sn_b, \end{equation}
using the properties of the four-dimensional spinors. Similarly, the contraction relation $\sk_a \sdot \sn_b = \epsilon_{ab}$ is upheld, as are the spinor transformations. Of course, the reality conditions are also unaffected. We can therefore proceed and use these results for curved spacetime.

\subsection{The little group spinors}\label{Section::GR set up}

In order to construct the Weyl spinor $\Psi_{ABCD}$, we simply contract the Weyl tensor $C_{\mu\nu\rho\sigma}$ with the curved space gamma basis to obtain
\begin{equation}\label{WeylSpinorDefn}
	\Psi_{ABCD}=C_{\mu\nu\rho\sigma}\,\sigma^{\mu\nu}{}_{AB}\,\sigma^{\rho\sigma}{}_{CD}
\end{equation}
as in section \ref{setupandcoarseclassfieldstrength}. The rotation generator $\sigma^{\mu\nu}{}_{AB}$ is of course constructed from the curved space $\gamma$'s now but is otherwise defined as in equation \eqref{RotationMatrix}. Given the symmetries of the Weyl tensor, it is easy to show that the Weyl spinor is totally symmetric, and thus comprises the 35 degrees of freedom in the five-dimensional Weyl tensor. 

As in section \ref{setupandcoarseclassfieldstrength}, we would like to break up these 35 degrees of freedom according to their boost weight by contracting in our (unchanged) spinor basis. The little group objects $\Psi^{(i)}_{abcd}$ are defined by 
\begin{equation}\label{littlegroupspinordefns}
\begin{split}
\Psi^{(0)}_{abcd}=&\;\Psi_{ABCD}\;\sk^A{}_a \, \sk^B{}_b \,\sk^C{}_c \,\sk^D{}_d 
\\\Psi^{(1)}_{abcd}=&\;\Psi_{ABCD}\;\sk^A{}_a\, \sk^B{}_b \,\sk^C{}_c \,\sn^D{}_d 
\\\Psi^{(2)}_{abcd}=&\;\Psi_{ABCD}\;\sk^A{}_a \,\sk^B{}_b \,\sn^C{}_c\, \sn^D{}_d 
\\\Psi^{(3)}_{abcd}=&\;\Psi_{ABCD}\;\sk^A{}_a \,\sn^B{}_b \,\sn^C{}_c\, \sn^D{}_d 
\\\Psi^{(4)}_{abcd}=&\;\Psi_{ABCD}\;\sn^A{}_a \,\sn^B{}_b \,\sn^C{}_c\, \sn^D{}_d,
\end{split}
\end{equation} where the bracketed superscript number $(i)$ indicates the number of $\sn^A{}_a$ spinors in the contraction. These definitions are analogous to the field strength objects $\Phi^{(i)}_{ab}$ in equation \eqref{phidefns} and to the four-dimensional definitions \eqref{4DGRdecomposition}. $\Psi_{ABCD}$ can equivalently be expressed as the sum of the little group objects:
\begin{equation}\label{Psidef}
\begin{split}
	\Psi_{ABCD}=&\;\Psi^{(0)}_{abcd}\; \sn_A{}^a \,\sn_B{}^b\, \sn_C{}^c\, \sn_D{}^d + 4 \,\Psi^{(1)}_{abcd} \;\sn_{(A}{}^a\, \sn_B{}^b \,\sn_C{}^c\, \sk_{D)}{}^d \\& +6\, \Psi^{(2)}_{abcd}\; \sn_{(A}{}^a \,\sn_B{}^b \,\sk_C{}^c \,\sk_{D)}{}^d
	\\& + 4\, \Psi^{(3)}_{abcd}\;\sn_{(A}{}^a \,\sk_B{}^b\, \sk_C{}^c \,\sk_{D)}{}^d + \Psi^{(4)}_{abcd}\; \sk_{A}{}^a\, \sk_B{}^b\, \sk_C{}^c \,\sk_{D}{}^d .
\end{split}
\end{equation}

We observe from the definitions of the little group objects $\Psi^{(i)}$ that they possess different symmetries. The totally symmetric ones, $\Psi^{(0)}_{abcd}$ and $\Psi^{(4)}_{abcd}$, have 5 degrees of freedom, while $\Psi^{(1)}_{abcd}=\Psi^{(1)}_{(abc)d}$ and $\Psi^{(3)}_{abcd}=\Psi^{(3)}_{a(bcd)}$ each contain 8. $\Psi^{(2)}_{abcd}=\Psi^{(2)}_{(ab)(cd)}$ comprises the final 9 degrees of freedom to reach 35. It is sensible to break these 4-spinors into irreducible respresentations of $SU(2)$. We will use the notation that a lower case $\psi^{(i)}$ indicates a totally symmetric object, i.e., $\psi^{(i)}_{abcd}=\psi^{(i)}_{(abcd)}$ for any value of $i$, and we also introduce $\chi^{(i)}$ to indicate a symmetric bi-spinor. Clearly $\Psi^{(0)}$ and $\Psi^{(4)}$ are already irreducible, since they sit in the totally symmetric representation $\textbf{5}$, so $\Psi^{(0)}=\psi^{(0)}$ and $\Psi^{(4)}=\psi^{(4)}$. $\Psi^{(1)}$ and $\Psi^{(3)}$ contain a bi-spinor trace that can be removed to decompose them as $\textbf{8}=\textbf{5}+\textbf{3}$:
\begin{equation}
\begin{split}
\Psi^{(1)}_{abcd} =&\; \psi^{(1)}_{abcd} - \frac{1}{4} \left(\epsilon_{ad}\chi^{(1)}_{bc}+\epsilon_{bd}\chi^{(1)}_{ac}+\epsilon_{cd}\chi^{(1)}_{ab}\right)
\\\Psi^{(3)}_{abcd} =&\; \psi^{(3)}_{abcd} - \frac{1}{4} \left(\epsilon_{ab}\chi^{(3)}_{cd}+\epsilon_{ac}\chi^{(3)}_{bd}+\epsilon_{ad}\chi^{(3)}_{bc}\right),\end{split}
\end{equation}
while $\Psi^{(2)}$ splits into a symmetric rank 4 spinor, a symmetric rank 2 spinor and a scalar: $\textbf{9}=\textbf{5}+\textbf{3}+\textbf{1}$ as
\begin{equation}
	\Psi^{(2)}_{abcd}= \psi^{(2)}_{abcd} - \frac{1}{4} \left(\epsilon_{ac}\chi^{(2)}_{bd}+\epsilon_{ad}\chi^{(2)}_{bc}+\epsilon_{bc}\chi^{(2)}_{ad}+\epsilon_{bd}\chi^{(2)}_{ac}\right)+\frac{1}{6}\left(\epsilon_{ac}\epsilon_{bd}+\epsilon_{ad}\epsilon_{bd}\right)\Psi^{(2)}_{\mathrm{tr}}.
\end{equation}This is summarised in table \ref{content breakdown}.
 \begin{table}[h]
 	\centering
 	\renewcommand{\arraystretch}{1.8}\begin{tabular}{|c||c||ccc||c|}
 		\hline
 		Reducible little group spinor &\hspace{30pt}&\hspace{10pt}\textbf{5}\hspace{10pt}&\hspace{10pt}\textbf{3}\hspace{10pt}&\hspace{10pt}\textbf{1}\hspace{10pt}&Total dof
 		\\\hline\hline
 		$\Psi^{{}(0)}_{abcd}=\Psi^{{}(0)}_{(abcd)}$&&\hspace{10pt}$\psi^{{}(0)}_{abcd}$&&&5
 		\\ $\Psi^{{}(1)}_{abcd}=\Psi^{{}(1)}_{(abc)d}$&&\hspace{10pt}$\psi^{{}(1)}_{abcd}$&\hspace{10pt}$\chi^{(1)}_{ab}$&&8
 		\\ $\Psi^{{}(2)}_{abcd}=\Psi^{{}(2)}_{(ab)(cd)}$&$\Rightarrow$& \hspace{10pt}$\psi^{{}(2)}_{abcd}$&\hspace{10pt}$\chi^{(2)}_{ab}$& \hspace{10pt}$\Psi^{(2)}_{\mathrm{tr}}$\hspace{10pt}&9
 		\\$\Psi^{{}(3)}_{abcd}=\Psi^{{}(3)}_{a(bcd)}$&&\hspace{10pt}$\psi^{(3)}_{abcd}$&\hspace{10pt}$\chi^{(3)}_{ab}$&&8
 		\\$\Psi^{{}(4)}_{abcd}=\Psi^{{}(4)}_{(abcd)}$&&\hspace{10pt}$\psi^{(4)}_{abcd}$&&&5
 		\\\hline
 	\end{tabular}
 	\caption{The table shows how each little group 4-spinor is decomposed into irreducible representations. $\textbf{5}$ is a totally symmetric 4-spinor, \textbf{3} is a symmetric bi-spinor, and \textbf{1} is a scalar. We write ``dof'' as a short-hand for degrees of freedom.}
 	\label{content breakdown}
 \end{table}

 We will also use vectorial language for the little group irreps, translating between the two using the Pauli matrices $\varsigma^i{}_{ab}$ as usual such that, for example,
\begin{equation}
	\psi^{(0)}_{ij} = \varsigma_i{}^{ab} \, \varsigma_j{}^{cd} \, \psi^{(0)}_{abcd}.
\end{equation} Table \ref{so3} summarises the notation. This is a simple matter of representation, and makes it easier to compare our results with the vectorial techniques used in the literature. In this notation, imposing the reality conditions is equivalent to the requirement that the objects are real.
 	\begin{table}[h]
 		\centering
 		\renewcommand\arraystretch{1.6}
 		\begin{tabular}{|ccc||c||ccc|}
 			\hline
 			4-spinor&2-spinor &scalar&& 3-matrix&3-vector&scalar
 			\\ 
 			\hline \hline
 			$\psi^{(0)}_{abcd}$&&&&$\psi^{(0)}_{ij}$&&
 			\\ $\psi^{(1)}_{abcd}$&$\chi^{(1)}_{ab}$&&& $\psi^{(1)}_{ij}$&$\underline{\chi}^{(1)}$&
 			\\$ \psi^{(2)}_{abcd}$&$\chi^{(2)}_{ab}$&$ \Psi^{(2)}_{\mathrm{tr}}$&$\hspace{10pt}\leftrightarrow\hspace{10pt}$&$\psi^{(2)}_{ij}$&$\underline{\chi}^{(2)}$&$\Psi^{(2)}_{\mathrm{tr}}$
 			\\
 			$\psi^{(3)}_{abcd}$&$\chi^{(3)}_{ab}$&&& $\psi^{(3)}_{ij}$ &$\underline{\chi}^{(3)}$&
 			\\$\psi^{(4)}_{abcd}$&&&&$\psi^{(4)}_{ij}$&&
 			\\\hline
 		\end{tabular}
 		\caption{The irreducible representations of the Weyl spinor can be easily moved between spinor space on the left and vector space on the right by use of the Pauli matrices $\varsigma^i{}_{ab}$. We will use the two notations interchangeably. Note that all spinors are totally symmetric, and that all 3-matrices are symmetric and tracefree.}
 		\label{so3}
 	\end{table}

\subsubsection{Coarse and finely grained classifications}\label{finegrain}

This construction naturally highlights two levels of classification, one coarse-grained which depends only on the little group spinors, and one which is more finely grained which also depends on the irreducible representation. The coarse classification arises due to the similarities in construction between the little group spinors
\begin{equation}
	\Psi^{(i)}_{abcd}, \quad\quad i=1,...,4,
\end{equation} defined in equation \eqref{littlegroupspinordefns}, and the complex scalars from four dimensions
\begin{equation}
	\psi_i \quad \quad \quad i=1,...,4,
\end{equation} defined in equation \eqref{4DGRdecomposition}. Thus the $\Psi^{(i)}$ will obey a classification which is analogous to the four-dimensional Petrov one shown in table \ref{4dpetrovweyl} \footnote{There is one caveat, which is that in four dimensions it is always possible to find a tetrad where $\psi_0$ vanishes. This is not the case in general so we require the additional type G to account for such spacetimes; see \cite{Coley2004a}.}. This coarse classification is proposed in table  \ref{petrov5d} and as we will show in section \ref{cmpp}, it turns out to be equivalent to the CMPP classification \cite{Coley2004a,Coley2004b}.

 \begin{table}[h]
 	\centering
 	\renewcommand{\arraystretch}{1.5}\begin{tabular}{|l l  |} 
 		\hline
 		Type & Little group spinor characteristic
 		\\ 
 		\hline
 		Type G &   $\Psi^{(0)}\neq0$  
 		\\ 
 		Type I &   $\Psi^{(0)}=0$  
 		\\ 
 		Type II &  $\Psi^{(0)}=\Psi^{(1)}=0$  
 		\\
 		Type D &  $\Psi^{(0)}=\Psi^{(1)}=\Psi^{(3)}=\Psi^{(4)}=0$  
 		\\ 
 		Type III &  $\Psi^{(0)}=\Psi^{(1)}=\Psi^{(2)}=0$  
 		\\
 		Type N &  $\Psi^{(0)}=\Psi^{(1)}=\Psi^{(2)}=\Psi^{(3)}=0$  
 		\\
 		\hline
 	\end{tabular}
 	\caption{Table showing the coarse grained, Petrov-like classification of a five-dimensional Weyl tensor built in analogy with the four-dimensional Petrov formalism. The classification refers to the vanishing of the reducible little group spinors $\Psi^{(i)}$, which is equivalent to the vanishing of a whole row in table \ref{content breakdown}.}
 	\label{petrov5d}
 \end{table}
 
The fine grained classification notes that the coarse types in table \ref{petrov5d} referred only to the rows of table \ref{content breakdown}. The columns spreading out into different irreducible representations of the little group shows that a greater level of detail is possible. For example, imagine two type D solutions: then a pentad can be found for each where only $\Psi^{(2)}$ is non-zero. Suppose further that when the fine structure is analysed, it is seen that $\chi^{(2)}$ and $\psi^{(2)}$ vanish for the first spacetime but only $\chi^{(2)}$ vanishes for the second, indicating that the first example is more special. This is exactly the case for the Tangherlini-Schwarzschild black hole and the black string respectively - the details of this example are given in the following section.

We can delve deeper into the irreps themselves to ask whether they also have sub-classifications. First we will consider a complex spacetime. In this case, the structure of the irreducible representations $\psi^{(i)}$ and $\chi^{(i)}$, namely complex symmetric spinors with two-dimensional indices, is exactly that of the four-dimensional Weyl and field strength spinors respectively. Like a Russian doll, hiding inside the Weyl tensor are additional lower-dimensional Weyl tensors. These also have a classification, which can be found in the usual way for four dimensions. For example, a 4-spinor $\psi_{abcd}=\alpha_{(a}\beta_b\gamma_c\delta_{d)}$ could have any of four different specialisations:
 \begin{itemize}
 	\item Type II: Two repeated spinors with the other two spinors distinct \newline $\psi_{abcd}~=~\alpha_{(a}\alpha_b\gamma_c\delta_{d)}$ 
 	\item Type D: Two pairs of repeated spinors  $\psi_{abcd}=\alpha_{(a}\alpha_b\gamma_c\gamma_{d)}$
 	\item Type III: Three repeated spinors $\psi_{abcd}=\alpha_{(a}\alpha_b\alpha_c\delta_{d)}$ 
 	\item Type N: Four repeated spinors $\psi_{abcd}=\alpha_{a}\alpha_b\alpha_c\alpha_d$ 
 \end{itemize}
whereas for a 2-spinor $\chi_{ab}=\alpha_{(a}\beta_{b)}$ there is only one specialisation
\begin{itemize}
	 	\item Type II: Two repeated spinors $\chi_{ab}=\alpha_{a}\alpha_b$ .
 \end{itemize}
 
 In contrast, when we restrict to a real spacetime we find that much of this second layer of hidden lower-dimensional Weyl tensor classification is forbidden. We already know from our analysis of the field strength tensor in section \ref{setupandcoarseclassfieldstrength} that a bi-spinor $\chi^{(i)}$ which obeys the reality conditions ${\chi}=\overline{\chi}$ cannot be written as the outer product of a single spinor, $\chi_{ab}\neq \alpha_a\alpha_b$. A similar analysis can be applied to real symmetric 4-spinor objects $\psi_{abcd}$ which satisfy $\psi=\overline{\psi}$. This will restrict the number of subclasses available, as we will now show.
 
It is well known from four dimensions (see for example \cite{Stewart:1990uf}) that if we define $I=\psi^{abcd}\, \psi_{abcd}$ and $J=\psi_{ab}{}^{cd}\,\psi_{cd}{}^{ef}\,\psi_{ef}{}^{ab}$, then the requirements for each class are:
 \begin{itemize}
 	\item Type II: $I^3=6J^2$
 	\item Type D: $\psi_{pqr(a}\,\psi_{bc}{}^{pq}\,\psi^r{}_{def)}=0$
 	\item Type III: $I=J=0$
 	\item Type N: $	\psi_{(ab}{}^{ef}\,\psi_{cd)ef}=0$.
 \end{itemize}
Since our $\psi$'s obey the reality condition, they can be rewritten as symmetric tracefree matrices with real entries. In contrast, if we had chosen to consider complex space, or a different signature, the entries would be complex. A real symmetric matrix may always be diagonalised to obtain
 \begin{equation}
 D=\begin{pmatrix}
 \lambda_1&0&0\\0&-\left(\lambda_1+\lambda_2\right)&0\\0&0&\lambda_2
 \end{pmatrix}
 \end{equation}
 and so we can rewrite the conditions in terms of the eigenvalues as
  \begin{itemize}
  	\item Type II: $2 \lambda_1^3+3 \lambda_1^2 \lambda_2-3 \lambda_1 \lambda_2^2-2 \lambda_2^3=0$
  	\item Type D: $2 \lambda_1^3+3 \lambda_1^2 \lambda_2-3 \lambda_1 \lambda_2^2-2 \lambda_2^3=0$
  	\item Type III: $\lambda_1^2+\lambda_1\lambda_2+\lambda_2^2=0$ and $\lambda_1\lambda_2(\lambda_1+\lambda_2)=0$
  	\item Type N: $	\lambda_1^2=\lambda_2^2$ and $\lambda_1^2+4\lambda_1\lambda_2+\lambda_2^2=0$.
  \end{itemize}
The type II condition has reduced to the more specialised type D condition and is solved only when two of the eigenvalues are equal (or trivially when all the eigenvalues vanish). In contrast, there are no non-trivial solutions for type N and type III, that is, we must have $\lambda_1=\lambda_2=0$. This tells us that under our reality conditions, only type D-like lower-dimensional Weyl tensors are possible.\footnote{We note that this argument is invalidated when complex entries occur because in general complex symmetric matrix cannot be diagonalised. } We note that interesting behaviour relating to dimensional reduction also occurs when a single eigenvalue vanishes, which is not reflected by this classification. We hope to explore this property further in future work.

To summarise, we have found three layers of structure naturally embedded in our formalism. The first is a Petrov-like coarse layer in the little group spinors. The second is more fine-grained, breaking the little group spinors into irreducible representations. Finally, the third looks at the irreps themselves and uses their similarity to four-dimensional objects to classify them in a Petrov-like way. This has two possibilities depending on whether or not reality conditions have been imposed as summarised in table \ref{Table::OnionStructure}.
\begin{table}[h!]
	\centering
	\begin{tabular}{c c l}
		\hline
Complex&$\psi$:& I, II, D, III, N
\\&$\chi$:& I, II
\\\hline
Real&$\psi$:&  I, D
\\&$\chi$:& I
\\\hline
	\end{tabular}
	\caption{The classification of the lower-dimensional objects hidden within the Weyl tensor depends on whether or not reality conditions have been imposed.}
	\label{Table::OnionStructure}
\end{table}

 \subsection{Examples}
 
 To illustrate a few key features of the formalism, we shall give a few very simple examples: the plane wave, a Tangherlini-Schwarzschild black hole and a black string.
 
 \subsubsection{A pp-wave}
 
 The metric for pp-wave can be expressed in Brinkmann coordinates 
 \begin{equation}
ds^2=-H(u,x,y,z) du^2 - 2 du\, dv +dx^2+dy^2+dz^2,
 \end{equation}
 such that if we choose the pentad 
 \begin{equation}
 	k= \partial_v , \quad n=\partial_u - \frac{1}{2}\, H(u,x,y,z) \partial_v ,\quad \varepsilon_{ab}=\begin{pmatrix}
 	\partial_x+ i\partial_y & i \partial_z \\ i \partial_z & 	\partial_x- i\partial_y 
 	\end{pmatrix} ,
 \end{equation}
  then the Weyl tensor is given by 
 \begin{equation}
 	C_{\mu\nu\rho\sigma} = 2\,\partial_i \partial_j H(u,x,y,z) \, n_{[\mu}\, \varepsilon^i_{\nu]}\, n_{[\rho}\, \varepsilon^j_{\sigma]},
 \end{equation}
where the index $i=1,2,3$ runs over the three polarisation directions $\{x,y,z\}$ as usual, and we recall \eqref{eq:5dpolvec}. Recasting this as a spinor using the curved space gamma basis we find
 \begin{equation}
 \begin{split}
 	\Psi_{ABCD}=&\; C_{\mu\nu\rho\sigma}\sigma^{\mu\nu}{}_{AB}\sigma^{\rho\sigma}{}_{CD}\\
 	=&\;4\, \partial_i \partial_j H(u,x,y,z) \, \varsigma^i{}_{ab}\,  \varsigma^j{}_{cd}\,  \sk_A{}^a\,  \sk_B{}^b\, \sk_C{}^c\,  \sk_D{}^d.
 	 \end{split}
 \end{equation}
 Therefore the pp-wave is a type N solution with $\psi^{(4)}_{ij}=4 \, \partial_i \partial_j H(u,x,y,z) $. If we were to specify the function $H(u,x,y,z)$ we could classify $\psi^{(4)}_{abcd}$ further since it has all of the properties of a four dimensional Weyl tensor.
 
 \subsubsection{The Tangherlini-Schwarzschild black hole}
 
 Another simple example is a five-dimensional Schwarzschild black hole, with metric
 \begin{equation}
ds^2 = -\Delta(r) du^2 - 2\, du \, dr +r^2\left( d\theta^2+\sin^2\theta \left(d\phi^2 + \sin^2\, \theta d\chi^2\right)\right),
\end{equation}
where $\Delta(r)=1-\frac{r_s^2}{r^2}$\,. We choose the pentad 
 \begin{equation}
 k=-\partial_u + \frac{1}{2}\, \Delta(r)\, \partial_r , \quad n= \partial_r ,\quad \varepsilon_{ab}=\frac{1}{r}\begin{pmatrix}
 \partial_\theta + i\csc\theta \, \partial_\phi \;& i \csc\theta\csc\phi\, \partial_\chi \\ i \csc\theta\csc\phi\, \partial_\chi \;& 	\partial_\theta- i\csc\theta \, \partial_\phi
 \end{pmatrix} ,
 \end{equation}
 such that the Weyl tensor is 
 \begin{equation}\label{TangherliniWeyl}
 	C_{\mu\nu\rho\sigma}=\frac{2r_s^2}{r^4}\left(
 	2 k_{[\mu}\, \varepsilon^i_{\nu]}\, n_{[\rho}\, \varepsilon^i_{\sigma]}+2n_{[\mu}\, \varepsilon^i_{\nu]}\, k_{[\rho}\, \varepsilon^i_{\sigma]} -6 k_{[\mu}\, n_{\nu]}\, k_{[\rho}\, n_{\sigma]} 
 	-
 \varepsilon^i_{[\mu}\, \varepsilon^j_{\nu]}
 \varepsilon^{[i}_{[\rho}\, \varepsilon^{j]}_{\sigma]}
 	\right).
 \end{equation}The Weyl spinor is
 \begin{equation}
 \begin{split}
 \Psi_{ABCD}=&\; C_{\mu\nu\rho\sigma}\sigma^{\mu\nu}{}_{AB}\sigma^{\rho\sigma}{}_{CD}\\
 =&\; -\frac{48r_s^2}{r^4}  \left(\epsilon_{ac}\epsilon_{bd}+\epsilon_{ad}\epsilon_{bc}\right) \sk_{(A}{}^a\,  \sk_B{}^b\, \sn_C{}^{c}\,  \sn_{D)}{}^d,
 \end{split}
 \end{equation}
and so we can read off that the only non-zero little group irrep for the Tangherlini-Schwarzschild black hole is the scalar  $\Psi^{(2)}_{\mathrm{tr}} =-\frac{48r_s^2}{r^4}  $. Therefore, it is a very special type D solution, since it only has a single non-zero irrep. 

 \subsubsection{The black string}
 
 It is interesting to contrast this with another type D solution, the black string. This is a four-dimensional Schwarzschild black hole trivially extended along the $x^4=z$ direction with the metric 
 \begin{equation}
 ds^2 =- \Gamma(r) du^2 - 2\, du \, dr +r^2\left( d\theta^2+\sin^2\theta d\phi^2 \right) + dz^2
 \end{equation}
  where $\Gamma(r) =1-\frac{r_s}{r}$\,. We choose a pentad which is similar to the previous example:
   \begin{equation}
    k= \partial_r , \quad n=\partial_u- \frac{1}{2}\, \Gamma(r)\, \partial_r,\quad \varepsilon_{ab}=\frac{1}{r}\begin{pmatrix}
   \partial_\theta + i\csc\theta \, \partial_\phi & i  \partial_z \\ i  \partial_z & 	\partial_\theta- i\csc\theta \, \partial_\phi
   \end{pmatrix} ,
   \end{equation}
   to find that the Weyl tensor is 
 \begin{equation}
 \begin{split}
 C_{\mu\nu\rho\sigma}=2\frac{r_s}{r^3}\big(
2 \, \delta_{\mathrm{red}}^{ij} \left( k_{[\mu}\, \varepsilon^i_{\nu]}\, n_{[\rho}\, \varepsilon^j_{\sigma]}+   n_{[\mu}\, \varepsilon^i_{\nu]}\, k_{[\rho}\,  \varepsilon^j_{\sigma]}\right)&\;-2 k_{[\mu}\, n_{\nu]}\, k_{[\rho}\, n_{\sigma]} \\&\;
+\delta_{\mathrm{red}}^{ik} \,\delta_{\mathrm{red}}^{jl}\,
 \varepsilon^i_{[\mu}\, \varepsilon^j_{\nu]}
 \varepsilon^{[k}_{[\rho}\, \varepsilon^{l]}_{\sigma]}
 \big),
 \end{split}
  \end{equation}
 where the reduced identity matrix $\delta_{\mathrm{red}} $ is trivial in the $z$ direction, $\delta_{\mathrm{red}}^{ij}=\delta^{ij}-e_z^ie_z^j$. 
 Note the similarity to equation \eqref{TangherliniWeyl} if $\delta_{\mathrm{red}}$ is replaced by $\delta$. As usual, we recast as a spinor to find
 \begin{equation}
 	\Psi_{ABCD} =  -\frac{96 r_s}{r^3}\, \delta_{\mathrm{red}}^{ij}\, \varsigma^i{}_{ab}\,\varsigma^j{}_{cd} \, \sk_{(A}{}^a\,  \sk_B{}^b\, \sn_C{}^{c}\,  \sn_{D)}{}^d.
 \end{equation}
 This time there is more than one little group irrep present. The reducible little group spinor $\Psi^{(2)}$ is given by
 \begin{equation}
 \begin{split}
 	\Psi^{(2)}{}^{ij}=-\frac{4 r_s}{r^3} \delta_\mathrm{red}^{ij},
  \end{split}
  \end{equation} which decomposes into a trace term and a traceless symmetric \textbf{5}:
  \begin{equation}
 \psi^{(2)}{}^{ij} = -\frac{4 r_s}{r^3} \left(\frac13\, \delta^{ij}- e_z^i e_z^j \right),   \quad\quad \Psi^{(2)}_{\mathrm{tr}}=-\frac{16 r_s}{r^3}.
  \end{equation}  Therefore the black string is still a type D solution but it has a very different fine structure to the Tangherlini-Schwarzschild black hole. 
  
Finally, we can consider the structure of $\psi^{(2)}$ itself: since it has two equal eigenvalues ($\lambda_x=\lambda_y=-\frac{4r_s}{3r^3}$), the irrep is itself type D. 
 
\subsection{Relations to the literature: CMPP and de Smet}

As we have previously mentioned, there exist previously proposed classifications for five dimensions. Two of these are the classification derived by Coley, Milson, Pravda and Pravdova (CMPP) in 2004 \cite{Coley2004a,Coley2004b} and the de Smet classification proposed in 2002 \cite{deSmet2002}.
These two classifications are in disagreement, since some spacetimes are algebraically special in CMPP but not in de Smet, and vice versa. Their relationship was first investigated by Godazgar in 2010 \cite{Godazgar2010}.

\subsubsection{The CMPP classification}\label{cmpp}

In their papers \cite{Coley2004a,Coley2004b}, CMPP observe that each component of the Weyl tensor in five dimensions has a boost weight when the pentad is rescaled by $\{k,\,n,\,m^{(i)}\} \rightarrow \{\rho\, k,\, \rho^{-1} \,n,\,m^{(i)}\}$ for some scalar $\rho$, where $i =2,3,4$. This boost weight is the power of $\rho$ by which the component of the Weyl tensor transforms. The independent components of the Weyl tensor have the following boost weights:
\begin{equation}
\renewcommand\arraystretch{1.2}\begin{array}{|c||c|c|c|c|c|}
\hline
\mathrm{Boost \hspace{3pt}weight} & 2 & 1&0&-1&-2 \\ \hline
\mathrm{Component} & C_{0i0j} & C_{010i}, C_{0ijk} &C_{0101}, C_{01ij}, C_{0i1j}, C_{ijkl} &C_{011i}, C_{1ijk} & C_{1i1j}\\\hline
\end{array}
\end{equation}
where the index 0 indicates a contraction with $k$, the index $1$ indicates a contraction with $n$, and a Roman index $i$ corresponds to the space-like direction $m^{(i)}$. Our usage of $k$ and $n$ is identical, while the CMPP polarisation directions $m^{(i)}$ can be chosen to correspond to our $\varepsilon^\mu{}_{i}$ as
\begin{equation}
	m^{\mu (i)}=\varsigma^i\,{}^{ab}\, \varepsilon^\mu{}_{ab}.
\end{equation}
The Weyl tensor components, combined by boost weight, are then organised into a classification which is shown in table \ref{Table::CMPPClassification}. This is valid in any dimension, and of course reduces to the Petrov classification in four dimensions. 
 \begin{table}[h]
 	\centering
 	\renewcommand{\arraystretch}{1.5}\begin{tabular}{|l l|} 
 		\hline
 		Type & Characteristic\\ 
 		\hline
 		Type G &   $C_{0i0j}\neq0$  \\ 
 		Type I &   $C_{0i0j}=0$  \\ 
 		Type II &  $C_{0i0j}=C_{010i}=C_{0ijk}=0$ \\
 		Type D &  $C_{0i0j}=C_{010i}=C_{0ijk}=C_{011i}=C_{1ijk}=C_{1i1j}=0$  
 		\\ Type III &  $C_{0i0j}=C_{010i}=C_{0ijk}=C_{0101}=C_{01ij}=C_{0i1j}=C_{ijkl}=0$  \\
 		Type N &  $C_{0i0j}=C_{010i}=C_{0ijk}=C_{0101}=C_{01ij}=C_{0i1j}=C_{ijkl}=C_{011i}=C_{1ijk}=0$  \\
 		\hline
 	\end{tabular}
 	\caption{The CMPP classification considers the vanishing of the components of the Weyl tensor in some pentad in order to specify a type. The more special the classification, the more components, grouped by boost weight, must vanish.}
 	\label{Table::CMPPClassification}
 \end{table}

The boost transformation is clearly identical to the boost that we have previously defined through spinor space as $\sk_A{}^a\rightarrow c \, \sk_A{}^a$, $\sn_A{}^a\rightarrow \frac1c \, \sn_A{}^a$. As shown in equation \eqref{eq:boostgr}, the effect on the pentad is identical when we identify $\rho=c^2$. We therefore expect to see a correlation between the components of the Weyl tensor and the little group 4-spinors. This turns out to be exactly the case. We can easily use the equations \eqref{WeylSpinorDefn}, \eqref{kndefns} and \eqref{polarisationdefn}, which express the Weyl tensor, $k$, $n$ and $\varepsilon^\mu{}_{ab}$ in terms of spinors, to show that the CMPP components correspond directly to little group irreps:
 \begin{equation}
 \setlength{\arraycolsep}{12pt}\renewcommand{\arraystretch}{2.2}\begin{array}{lll}\label{cmpptospinordictionary}
 C_{0i0j} = \frac18 \psi^{(0)}_{ij} & C_{010i}=-\frac1{8\sqrt{2}} \chi^{(1)}_{i}
 &C_{0ijk} =\frac1{8\sqrt{2}}\left( 2\, \epsilon_{ijl}\,  \psi^{(1)}_{lk}-\chi^{(1)}_{[i}\delta_{j]k}\right)
 \\
C_{0101}=\frac{1}{16} \Psi^{(2)}_{\mathrm{tr}}
 & C_{01ij}=- \frac1{8}\, \epsilon_{ijk}\chi^{(2)}_k
  &C_{0i1j}=-\frac{1}{8}\left( \psi^{(2)}_{ij}+\frac12 \epsilon_{ijk}\chi^{(2)}_{k} +\frac16 \Psi^{(2)}_{\mathrm{tr}} \delta_{ij}\right)
 \\
 C_{1i1j}= \frac18 \psi^{(4)}_{ij}
 &C_{011i}=\frac1{8\sqrt{2}}\chi^{(3)}_{i}
 &C_{1ijk} = -\frac1{8\sqrt{2}}\left(2\epsilon_{ijl}\psi^{(3)}_{lk}+\chi^{(3)}_{[i}\delta_{j]k}\right)
 \\&\multicolumn{2}{l}{C_{ijkl} =\frac{1}{2} \left(\delta_{i[l}\,\psi^{(2)}_{k]j}-\delta_{j[l}\,\psi^{(2)}_{k]i}+\frac1{12}\Psi^{(2)}_{\mathrm{tr}}\,\delta_{i[l}\,\delta_{k]j}\right).} 
 \end{array}
 \end{equation}
Using this correspondence, it is clear that the classifications shown in tables \ref{Table::CMPPClassification} and \ref{petrov5d} are identical. Thus, the coarse classification inspired by the similarities of our construction with the four-dimensional Petrov classification is exactly the original CMPP classification.

\subsubsection{Little group irreps }

The irreducible representations $\psi^{(i)}$, $\chi^{(i)}$ and $\Psi^{(2)}_{\mathrm{tr}}$ also make an appearance in the literature. It was noted in \cite{Coley2009} that there are subgroups of the Weyl components for a given boost weight by noting their grouping under Lorentz transformations. For example, Coley and Hervik define two subclasses of type I by
\begin{itemize}
\item Type I(A) $\Leftrightarrow
C^i{}_{ji0}=0$
\item Type I(B) $\Leftrightarrow
C_{ijk0} \, C^{ijk}{}_0=\frac12 C^{ji}{}_{j0} \, C^{k}{}_{ik0}$
\end{itemize} in the Weyl-aligned basis for an arbitrary number of dimensions. As before, we can cast this into little group space in five dimensions to find that this corresponds to
\begin{itemize}
\item Type I(A) $\Leftrightarrow
\chi^{(1)}_{ab}=0$
\item Type I(B) $\Leftrightarrow \psi^{(1)}_{abcd} =0$.
\end{itemize} The other little group irreps are identified in a similar way. In \cite{Coley2012}, now joined by Ortaggio and Wylleman, Coley and Hervik apply their results to five dimensions and find that the Weyl tensor can be written in terms of 5 symmetric trace-free  matrices, three vectors and a scalar, which produce exactly the fine structure that we presented based on spinor-helicity considerations. Thus, the spinorial techniques we have developed are precisely the spinor underpinnings of the refined CMPP classification.

\subsubsection{The de Smet classification}\label{Section::deSmet}
\begin{figure}[h]
	 	\centering
	 	\includegraphics[width=.9\linewidth]{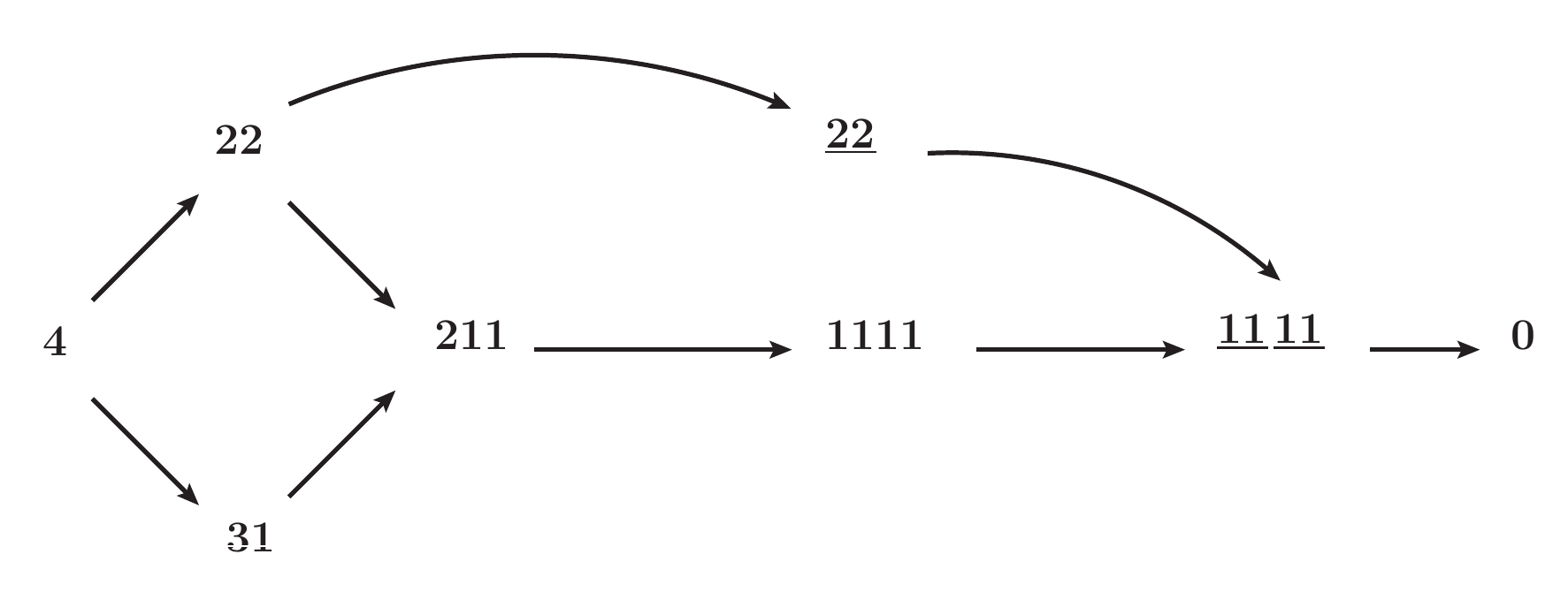}
	 	\caption{The real de Smet classification proposed by \cite{deSmet2002} and restricted with reality conditions by \cite{Godazgar2010} contains 8 classes including the flat spacetime class $\textbf{0}$, for which the Weyl tensor vanishes.}
	 	\label{figure::deSmetClassification}
	 \end{figure}

As we previously mentioned, another notable higher-dimensional classification is that of de Smet \cite{deSmet2002}. In this work, de Smet constructs the $SO(4,1)$ 4-spinor $\Psi_{ABCD}$ exactly as we have done, and then constructs a classification based on the factorisation properties of the Weyl polynomial $\mathcal{W}$, defined by
\begin{equation}
\mathcal{W}
\equiv \Psi_{ABCD}\,\xi^A \,\xi^B\, \xi^C\,\xi^D,
\end{equation}
for an arbitrary $\xi^A$. Originally containing 12 classes, further work by Godazgar \cite{Godazgar2010} found that consideration of the reality conditions brought the total number of classes down to 8. It was proposed that these can be  arranged in order of ``specialness'' as shown in figure \ref{figure::deSmetClassification}. We only consider real spacetimes in this section. The de Smet labels work as follows. The numbers indicate the rank of each factorised part of the Weyl polynomial and groups of underlined numbers signify that these are repeated factors. Thus, a \textbf{211}  indicates a Weyl polynomial with one factor quadratic in $\xi$ and two factors linear in $\xi$. If the spacetime is a \textbf{\underline{22}}, then there are two identical quadratic factors.
	 
We can interpret the de Smet construction in terms of our formalism by expanding equation \eqref{Psidef} in terms of its little group irreps. Because our formalism splits the spacetime into totally symmetric little group irreps, the factorisation properties can be easily investigated. To take a simple example, let us consider a spacetime for which only $\psi^{(2)}_{\mathrm{tr}}$ is non-zero (such as the Schwarzschild-Tangherlini solution), so that
\begin{equation}
\begin{split}
\mathcal{W}=&\;  \psi^{(2)}_{\mathrm{tr}} (\epsilon_{ac}\, \epsilon_{bd}+\epsilon_{ad}\, \epsilon_{bc})\, (\sn\sdot\xi)^a\, (\sn\sdot\xi)^b\,  (\sk\sdot\xi)^c \, (\sk\sdot\xi)^d
\\=&\; 2 \, \psi^{(2)}_{\mathrm{tr}}\,  [(\sn\sdot\xi),(\sk\sdot\xi)]^2.
\end{split}
\end{equation}
We have used $[\cdot,\cdot]$ to indicate a contraction on little group spinor indices, distinguishing it from the centre dot ``$\;\sdot\;$'' used to indicate contraction on spacetime spinor indices. Clearly, this factorises beautifully into a de Smet $\textbf{\underline{22}}$, which means that the Weyl polynomial factorises into two identical bi-spinors. 

Next, consider a type III solution for which only $\chi^{(3)}$ is non-zero. The Weyl polynomial is
\begin{equation}
\begin{split}
\mathcal{W}=&\;  -\left(\epsilon_{ab}\, \chi^{(3)}_{cd}+\epsilon_{ac}\, \chi^{(3)}_{bd}+\epsilon_{ad}\, \chi^{(3)}_{bc}\right)\, (\sn\sdot\xi)^a \, (\sk\sdot\xi)^b \, (\sk\sdot\xi)^c\,  (\sk\sdot\xi)^d 
\\=&\; -3 \left[\sn\sdot\xi,\sk\sdot\xi\right] \left[\sk\sdot\xi,\theta^{(3)}\right]\left[\sk\sdot\xi,\kappa^{(3)}\right]  
,\end{split}
\end{equation}
where, in the last line, we have used the property that symmetric $SU(2)$ bi-spinors can always be written as the symmetrisation of two spinors to define $\chi^{(3)}_{ab}\equiv \theta^{(3)}_{(a}\,\kappa^{(3)}_{b)}$. This has de Smet type $\textbf{211}$. Using the $\sk\leftrightarrow \sn$ symmetry, we can see that $\chi^{(1)}$ must also be a \textbf{211}:
\begin{equation}
\begin{split}
\mathcal{W} =-3 \left[ \sn\sdot\xi, \sk\sdot\xi \right]\left[\sn\sdot\xi ,\theta^{(1)}\right]\left[\sn\sdot\xi,\kappa^{(1)}\right],
\end{split}
\end{equation}
where again we have defined $\chi^{(1)}_{ab}\equiv \theta^{(1)}_{(a}\, \kappa^{(1)}_{b)}$. By contrast, when $\chi^{(2)}$ gives the sole contribution to $\Psi_{ABCD}$, the Weyl polynomial has de Smet class $\textbf{22}$:
\begin{equation}
\begin{split}
\mathcal{W}=-3  \left[\sn\sdot\xi,k\sdot\xi\right] \left\{
\left[\sn\sdot\xi ,\theta^{(2)} \right]\left[\sk\sdot\xi,\kappa^{(2)}\right]+\left[\sn\sdot\xi,\kappa^{(2)}\right] \left[\sk\sdot\xi,\theta^{(2)}\right]
\right\}.
\end{split}
\end{equation}

The $\psi^{(i)}$'s also have characteristic de Smet types. For example, if only $\psi^{(4)}$ is non-zero as for a type N spacetime, then the Weyl spinor is oriented in the $k$ direction as
 \begin{equation}
\Psi_{ABCD}=\psi^{(4)}_{abcd}	\, \sk_{(A}{}^a\, \sk_B{}^b \,\sk_C{}^c \, \sk_{D)}{}^d.
\end{equation}   
The explicit symmetrisation on the little group indices is not required, and thus the Weyl polynomial factorises totally to form a de Smet $\textbf{1111}$:
\begin{equation}
\mathcal{W}=\left[\sk\sdot\xi,\alpha^{(4)}\right]\left[\sk\sdot\xi,\beta^{(4)}\right]\left[\sk\sdot\xi ,\gamma^{(4)}\right]\left[\sk\sdot\xi,\delta^{(4)}\right].
\end{equation}
Using the invariance of de Smet classes under the interchange $n\leftrightarrow k$, we can see that $\psi^{(0)}$ is also of this type. However, the remaining $\psi^{(i)}$ do require proper symmetrisation over the little group indices, leading to sums over the different permutations which do not factorise at all and are de Smet $\textbf{4}$'s. For example, the Weyl polynomial for $\psi^{(1)}$ is: 
 \begin{equation}
 \begin{split}
\mathcal{W}=&\; \sum\limits_{\mathrm{Perms}\;\{\alpha,\beta,\gamma,\delta\}} \left[\sk\sdot\xi,\alpha^{(1)}\right]\left[\sn\sdot\xi,\beta^{(1)}\right]\left[\sn\sdot\xi,\gamma^{(1)}\right]\left[\sn\sdot\xi,\delta^{(1)}\right]
 .
 \end{split}
\end{equation}
As usual, $\psi^{(3)}$ can be obtained by $k\leftrightarrow n$ interchange. The expression for $\psi^{(2)}$ is very similar, except that it contains 6 terms due to the symmetrisation over two $k$'s and two $n$'s.

 As we can see, the de Smet classification is highly sensitive to the fine structure of the Weyl tensor. This is summarised in table \ref{Table::deSmetvLittleGroup}. At this point, it is possible to see that the hierarchy between de Smet classes proposed in \cite{deSmet2002} and shown in figure \ref{figure::deSmetClassification} is not actually present. For example, the $\textbf{211}$ class does not contain the full $\textbf{1111}$ class. A spacetime formed of more than one irrep will generically be a de Smet $\textbf{4}$. Although some special multi-irrep spacetimes exist, which are detailed in appendix~\ref{CompositeSpacetimes}, there are not very many of them and they arise only in highly specialised circumstances. This explains the disagreement between the de Smet and CMPP classifications elucidated by Godazgar in \cite{Godazgar2010}. On the one hand, because the CMPP classification is sensitive to the presence of the reducible little group spinors, it attributes the same Petrov class to a number of different possible de Smet classes\footnote{Although of course the refined CMPP classification in \cite{Coley2009,Coley2012} captures the little group irreps in full detail.}. On the other hand, the de Smet classification is most sensitive to the presence of a single irrep, irrespective of its boost weight. The two classifications clearly disagree in the notion of algebraic specialness.

\begin{table}[h]
	\centering
\setlength{\tabcolsep}{15pt}\renewcommand{\arraystretch}{1.6}
\begin{tabular}{|ccc||c||ccc|}\hline
\multicolumn{3}{|c||}{Little group irreps}&&\multicolumn{3}{|c||}{de Smet class}\\\hline
 \hline
 $\psi^{\mathrm{}(0)}_{abcd}$&&&&$\textbf{1111}$&&
 \\$\psi^{\textbf{}(1)}_{abcd}$&$\chi^{(1)}_{ab}$&&& \textbf{4}&\textbf{211}&
 \\$ \psi^{\mathrm{}(2)}_{abcd}$&$\chi^{(2)}_{ab}$&$ \Psi^{(2)}_{\mathrm{tr}}$&$\leftrightarrow$&\textbf{4}&\textbf{22}&\underline{\textbf{22}}
 \\$\psi^{\mathrm{}(3)}_{abcd}$&$\chi^{(3)}_{ab}$&&&\textbf{4}&\textbf{211}&
 \\$\psi^{\mathrm{}(4)}_{abcd}$&&&&\textbf{1111}&&
 \\\hline
 \end{tabular}
 \caption{The de Smet class of each little group irrep. The irreps are arranged by boost weight in the vertical direction and by dimension in the horizontal direction. Note the reflection symmetry in the central horizontal line, indicating invariance under the $k\leftrightarrow n$ interchange.}
 \label{Table::deSmetvLittleGroup}
 \end{table}

\subsection{Further refinements}
\label{sec:spinorAlign}

The classification we propose is based on identifying representations of the little group: the $\psi^{(i)}_{abcd}$, for $i=0, \ldots, 4$,  $\chi^{(j)}_{ab}$, for $j=1, 2, 3$, and $\Psi^{(2)}_\mathrm{tr}$. An algebraically general spacetime has a full set of these objects, none of which are vanishing, and furthermore satisfying no algebraic relations amongst them. 

Algebraically special cases can occur in a number of ways. We have already observed that it is possible for some of the little group objects to vanish, and a more subtle possibility is that one or more of the $\psi^{(i)}_{abcd}$'s could be type D. In terms of spinors, we can always find two-component spinors $\alpha_a$, $\beta_b$, $\gamma_c$ and $\delta_d$ such that $\psi^{(i)}_{abcd} = \alpha_{(a} \beta_b \gamma_c \delta_{d)}$ for a particular $i$. In the type D case, there are really only two different spinors up to scaling. In group theoretic terms, this particular $\psi^{(i)}$ is actually a three-dimensional representation rather than a five-dimensional representation.

It is also possible to have situations in which spinors are shared among different little group objects. In the complex case, there are many possibilities, but in the real case we are more limited. It is still possible that $\chi^{(i)} \propto \chi^{(j)}$ for some choices of $i$ and $j$. Alternatively, it could happen that a particular $\psi$ could be composed of some $\chi$: e.g., $\psi^{(1)}_{abcd} = \chi^{(2)}_{(ab} \chi^{(2)}_{cd)}$. The de Smet classification can be sensitive to such alignments in particular cases, as we discuss in Appendix~\ref{CompositeSpacetimes}.

\section{Higher dimensions}
\label{highdim}

Although we focused on five dimensions, our approach is quite general. Indeed, our starting point, the spinor-helicity method, is available in any number of dimensions~\cite{Cheung2009,CaronHuot:2010rj,Boels:2012ie}. In this section we will briefly discuss the classification in six dimensions. As this is an even number of dimensions, we choose a chiral basis of spinors, with Clifford algebra
\begin{equation}
\sigma^\mu{}_{AB} \tilde \sigma^{BC \, \nu} + \sigma^\nu{}_{AB} \tilde \sigma^{BC\,\mu} = -2 \eta^{\mu \nu} \mathbbm{1}_A^C.
\end{equation}
It happens that the Lie algebra of the Lorentz group in six dimensions, $\mathfrak{so}(6)$, is isomorphic to $\mathfrak{su}(4)$. This is reflected in the facts that the spinor representation of $\mathfrak{so}(6)$ is the four-dimensional fundamental representation of $\mathfrak{su}(4)$. From the point of view of $\mathfrak{su}(4)$, the six-dimensional vector representation of $\mathfrak{so}(6)$ is the antisymmetric tensor product of two $\mathbf{4}$s. Consequently, we can choose $\sigma^\mu$ and $\tilde \sigma^\mu$ to be antisymmetric $4\times4$ matrices.

In six dimensions, the little group is $SO(4)\cong SU(2)\times SU(2)\, / \, \mathbb{Z}_2$, so our first task is to understand how this product group structure is encoded in the spinors. Let $k^\mu$ be a six-dimensional null vector; then we define spinors associated with the vector by
\begin{equation}
k \cdot \sigma_{AB} \sk^B{}^a = 0.
\end{equation}
The index $a$ labels linearly independent solutions of this equation. The matrix $k \cdot \sigma_{AB}$ has vanishing determinant and, in fact, has rank 2. Thus the label $a$ takes values 1 and 2.

How can we reconstruct the null vector $k$ from the spinor $\sk^A{}^a$? The observation that the $\mathbf{6}$ is an antisymmetric combination of two $\mathbf{4}$s is helpful. There are six linearly independent $4\times4$ antisymmetric matrices, so if we expand an antisymmetric combination of the two spinors $\sk^A{}^a$ (for $a = 1, 2$) on the basis $\sigma^\mu{}_{AB}$, the result is guaranteed to transform as a vector. Since $k^\mu$ is the only vector available, we simply have to fix the normalisation. Indeed,
\begin{equation}
k^\mu = \frac1{2\sqrt{2}}\,\sk^A{}_a\,\sigma^\mu{}_{AB}\, \sk^B{}^a,
\label{eq:6dvectorSplit}
\end{equation}
where $\sk^A{}_a = \epsilon_{ab} \sk^A{}^b$; from this perspective, the matrix $\epsilon_{ab}$ is introduced to antisymmetrise the two possible $\sk^a$ spinors. 

This expression, equation~\eqref{eq:6dvectorSplit}, is manifestly invariant under an $SU(2)$ transformation $\sk^a \rightarrow U^a{}_b \sk^b$. This is part of the $SO(4)$ little group. The other $SU(2)$ factor acts on the antichiral spinors defined via
\begin{equation}
k \cdot \tilde \sigma^{AB} \tilde \sk_B{}^{\dot a} = 0,
\end{equation}
which implies that we may also write $k^\mu$ as
\begin{equation}
k^\mu = \frac1{2\sqrt{2}}\,\tilde \sk_A{}_{\dot{a}}\,\tilde\sigma^\mu{}^{AB} \,\tilde \sk_B{}^{\dot{a}}.
\end{equation}

To construct the analogue of the NP tetrad in six dimensions we pick a second null vector $n$ with the property that $k \cdot n = - 1$, and introduce spinors $\sn^A{}^{\dot a}$ and $\tilde \sn_A{}^a$. Then
\begin{align}
n^\mu =&\; \frac1{2\sqrt{2}}\,\sn^A{}_{\dot{a}}\,\sigma^\mu{}_{AB}\, \sn^B{}^{\dot{a}} \\
=&\; \frac1{2\sqrt{2}}\,\tilde \sn_A{}_{{a}}\,\tilde\sigma^\mu{}^{AB} \,\tilde \sn_B{}^{{a}}.
\end{align}
The set of spinors $\sk^A{}^a$, $\sn^A{}^{\dot a}$, $\tilde \sk_A{}^a$, $\tilde \sn_A{}^a$ spans the spinor spaces, so it is a simple matter to break the 15 degrees of freedom of the 2-form spinor $F^A{}_B$ and the 84 degrees of freedom in the Weyl spinor $C^{AB}{}_{CD}$ into little group irreps. Because this is done in exactly the same way as we did for five dimensions (subject to the details of the spinor spaces), we are guaranteed that the connection to CMPP will continue to be expressed. The representations of the little group spinors are now labelled by two numbers in six dimensions, $(i,j)$, and the boost weight is given by their average. The CMPP classification is simply the statement that each row of tables \ref{Table::6DFieldstrength} and \ref{Table::6DWeylSpinor} for the 2-form and Weyl tensor, respectively, vanishes appropriately. 

\begin{figure}[h!]
	 	\centering
	 	\begin{subfigure}{0.62\textwidth}
\includegraphics[width=.9\linewidth]{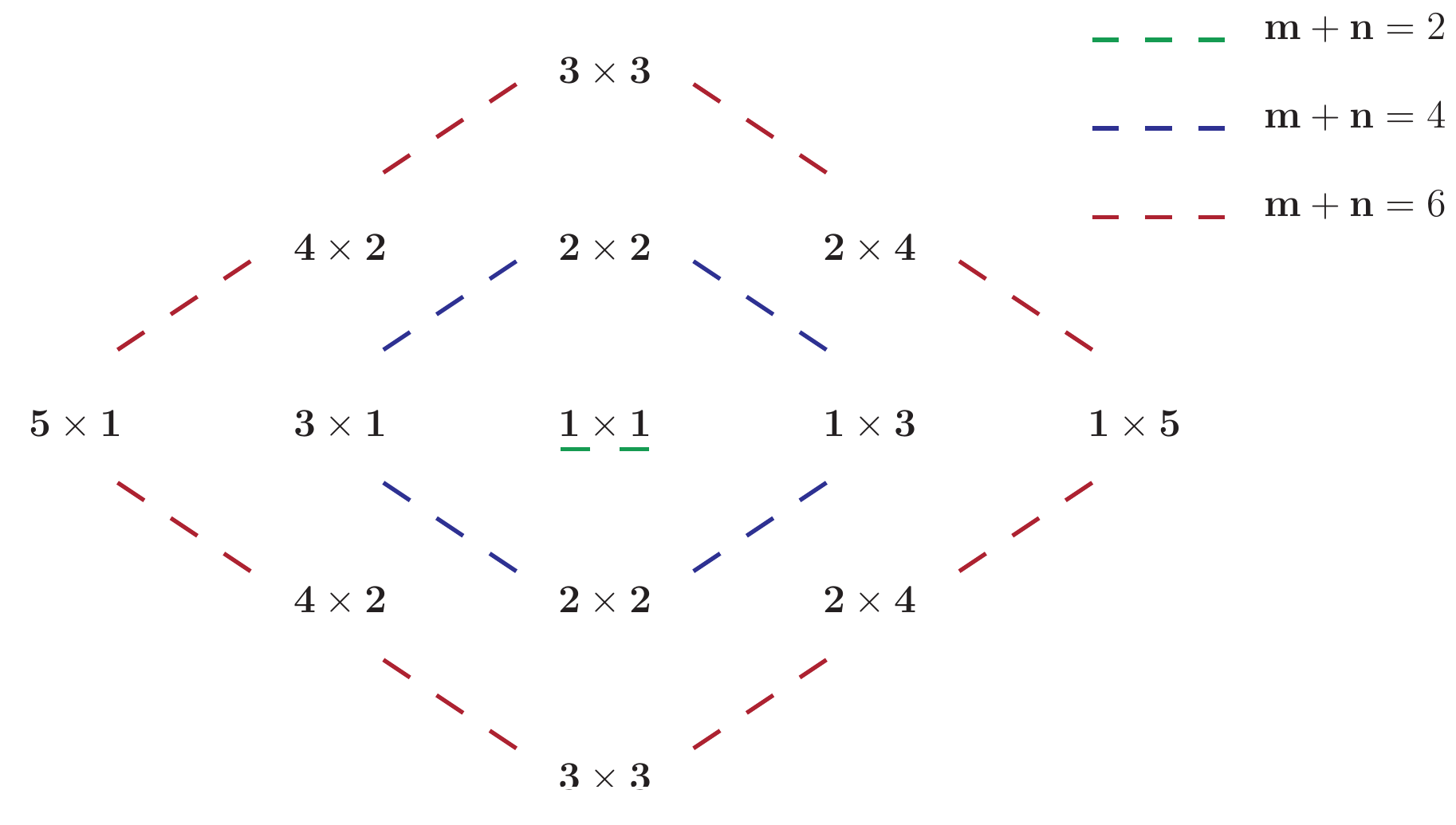}
	 	\end{subfigure}
	 \begin{subfigure}{0.37\textwidth}
	 		 		\includegraphics[width=.9\linewidth]{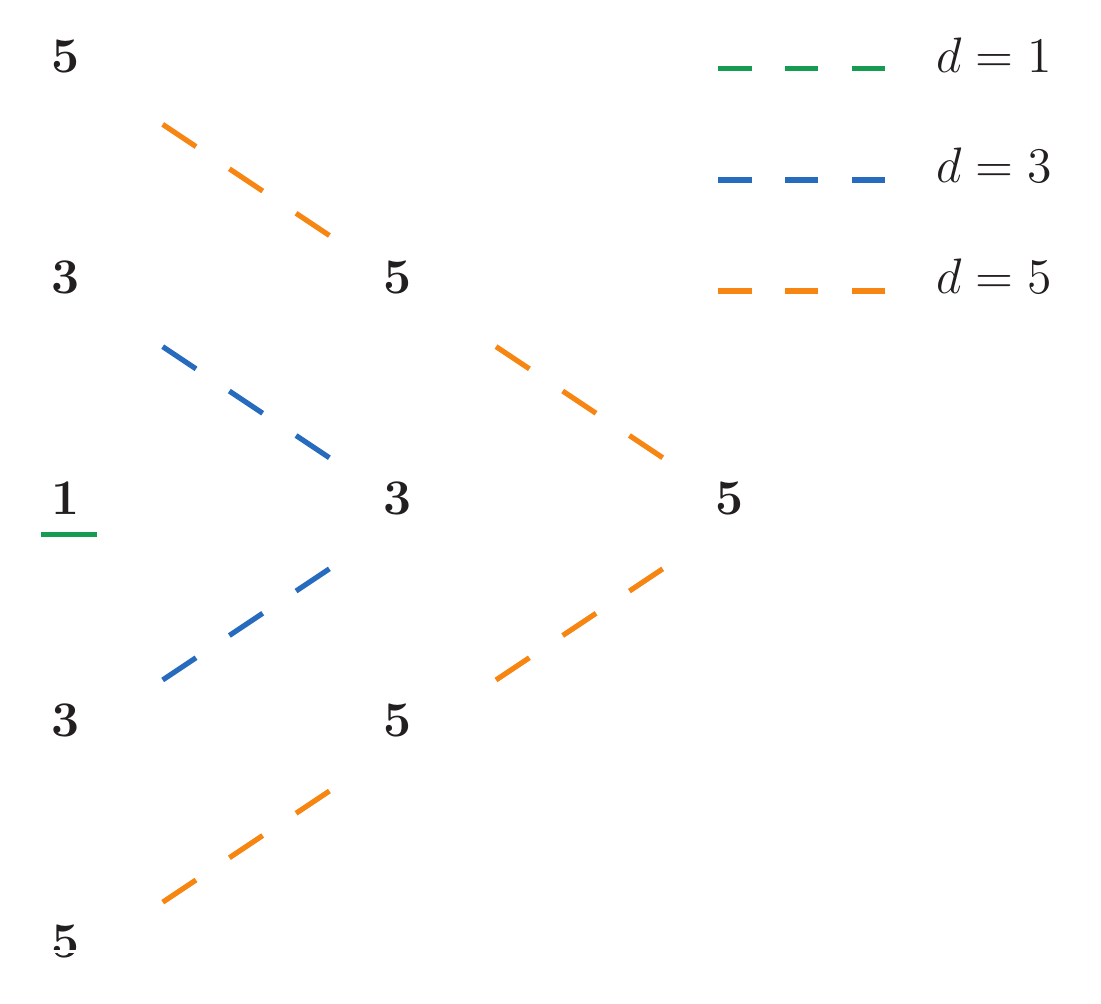}
		\end{subfigure}	 	
	 	\caption{The irreps of the six-dimensional Weyl spinor $\underline{m}\times\underline{n}$ form a kite-like pattern (left). Rows correspond to boost weight. Each concentric rhombus corresponds to a different value of $m+n$. Travelling clockwise from the leftmost value of the outer (red) rhombus, the values of $m$ decrease from 5 to 1 before increasing again, while $n$ increases from 1. A similar pattern is observed for the inner rhombus between 1 and 3. The irreps of the five-dimensional Weyl spinor (right) can be arranged in a similar way to six dimensions to form an arrowhead with concentric arrows of irrep dimension $d$. As usual, rows correspond to boost weight. }
	 	\label{figure::6DIrrepPattern}
	 \end{figure}

The appearance of a second number in the boost weight is due to a second symmetry in the irreps, that of an interchange between the two $SU(2)$ parts of the little group. This corresponds to an interchange $i\leftrightarrow j$ and dotted to undotted indices $a\leftrightarrow\dot{a}$, and manifests itself as a vertical line of symmetry through the centre of tables \ref{Table::6DFieldstrength} and \ref{Table::6DWeylSpinor}. This also explains the shape of the tables: previously, in five dimensions, where there was only a single $SU(2)$ little group, these decompositions had the shape of arrowheads which when reflected through the vertical axis form the characteristic rhombi of six dimensions. The dimensions of the irreps are not as regular as five dimensions, but have the pleasing distribution shown in figure \ref{figure::6DIrrepPattern} for the case of the Weyl spinor, laid next to their five-dimensional equivalent for comparison.

 \begin{table}[h!]
 	\centering
\setlength{\tabcolsep}{6pt} 	\renewcommand{\arraystretch}{1.8}\begin{tabular}{|c||c||c||c||c|}
 		\hline
 		Reducible spinors &&Irreducible spinors&&Irrep dimensionality\\
\hline\hline
 		 $\Phi^{{}(0,0)}_{a\dot{b}}$&&\hspace{10pt}$\phi^{(0,0)}_{a\dot{b}}$&&$\textbf{2}\times \textbf{2}$
 		\\ $\Phi^{{}(0,2)}_{ab}\hspace{60pt}\Phi^{(2,0)}_{\dot{a}\dot{b}}$&$\Rightarrow$&$\phi^{{}(0,2)}_{ab}\hspace{20pt}\Phi^{(1,1)}_{\mathrm{tr}}\hspace{20pt}\phi^{(2,0)}_{\dot{a}\dot{b}}$&$\Leftrightarrow$& $\textbf{1}\times \textbf{3} \hspace{10pt}\textbf{1}\times \textbf{1} \hspace{10pt} \textbf{3}\times \textbf{1} $
 		\\$\Phi^{(2,2)}_{\dot{a}b}$&&\hspace{10pt}$\phi^{(2,2)}_{\dot{a}b}$&&$\textbf{2}\times \textbf{2} $
 		\\\hline
 	\end{tabular}
 	\caption{The six-dimensional 2-form contains 4 reducible little group representations, which can be broken into 5 irreps. The rows are organised by boost weight, equal to the average of the bracketed superscripts. The columns are arranged such that the representations respect the $SU(2)$ interchange symmetry through the central vertical axis, hence the scalar $\Phi^{(1,1)}_{\mathrm{tr}}=\epsilon^{ab}\,\Phi^{(0,2)}_{ab}=\epsilon^{\dot{a}\dot{b}} \, \Phi^{(2,0)}_{\dot{a}\dot{b}}$ sits at the centre of the array.}
 	\label{Table::6DFieldstrength}
 \end{table}
 
\begin{table}[h!]
 	\centering
\setlength{\tabcolsep}{0pt}\renewcommand{\arraystretch}{2.5} \begin{tabular}{|ccccc||c||ccccc|}
		\hline
 \multicolumn{5}{|c||}{Reducible 6D little group spinors} && \multicolumn{5}{|c|}{Irreducible 6D little group spinors} \\	\hline\hline
 &&$\Psi^{(0,0)}_{(ab) \,(\dot{c}\dot{d})}$&&
 &&
 &&$\psi^{(0,0)}_{ab \, \dot{c}\dot{d}}$&&
 \\
 &$\Psi^{(0,2)}_{(ab)\,c \dot{d}}$&&$\Psi^{(2,0)}_{a\dot{b}\,\dot{c}\dot{d}}$&
 &&
 &$\psi^{(0,2)}_{abc \dot{d}}$&$\chi^{(1,1)}_{a\dot{b}}$&$\psi^{(2,0)}_{a\dot{b}\dot{c}\dot{d}}$&
  \\ \;$\Psi^{(0,4)}_{(ab)\,(cd)}$&&$\Psi^{(2,2)}_{a\dot{b}\, c\dot{d}}$&&$\Psi^{(4,0)}_{(\dot{a}\dot{b})\,(\dot{c}\dot{d})}$\;\;
  &
\;  \;$\Rightarrow$\;\;
  &
\;$\psi^{(0,4)}_{abcd}$\quad&\quad$\chi^{(1,3)}_{ab}$\quad&\quad$\Psi^{(2,2)}_{\mathrm{tr}}$\quad&\quad$\chi^{(3,1)}_{\dot{a}\dot{b}}$\quad&\quad$\psi^{(4,0)}_{\dot{a}\dot{b}\dot{c}\dot{d}}$\;\;
   \\
   &$\Psi^{(2,4)}_{\dot{a}b\,(c d)}$&&$\Psi^{(4,2)}_{(\dot{a}\dot{b})\,\dot{c}{d}}$&
   &&
   &$\psi^{(2,4)}_{\dot{a}bc {d}}$&$\chi^{(3,3)}_{\dot{a}b}$&$\psi^{(4,2)}_{\dot{a}\dot{b}\dot{c}{d}}$&
   \\
   &&$\Psi^{(4,4)}_{(\dot{a}\dot{b})\,(cd)}$&&
   &&
   &&$\psi^{(4,4)}_{(\dot{a}\dot{b})cd}$&& \\\hline
 \end{tabular}
 \caption{Connections between the traces of the reducible six-dimensional little group spinors allow us to break down the components into irreps. The indices of the reducible spinors (left) are organised in symmetrised pairs such that two like indices, for example $ab$ or $\dot{c}\dot{d}$ comprise 3 degrees of freedom each, while pairs such as $a\dot{b}$ and $\dot{c}d$ have no symmetrisation and constitute 4 degrees of freedom. For the table of irreducible representations on the right, all indices of the same $SU(2)$ type (ie dotted or undotted) are totally symmetric. The boost weight of each representation $(i,j)$ is given by $(i+j)/2$.}
 \label{Table::6DWeylSpinor}
 \end{table}

\section{Conclusions}
\label{conclusion}

We have demonstrated that higher-dimensional spinors provide a convenient formalism for the algebraic classification of spacetimes, extending Penrose's spinorial approach to the Petrov classification in four dimensions. The crucial element of the higher-dimensional spinorial construction, first proposed in \cite{Cheung2009} in the context of particle physics, is the explicit consideration of the little group. We have shown that the formalism not only leads naturally to the CMPP classification and its refinements, but it also allows for a natural connection with the de Smet classification. In particular, we have demonstrated that the de Smet classes mostly correspond to spacetimes where a single little group irrep is present, except for interesting cases where algebraic relations exist between distinct irreps. This analysis completes the work begun by \cite{Godazgar2010}. 

In this work, we have set up a basic framework but there is much to be done. We have not described in detail the choice of vector basis (pentad in five dimensions) that makes manifest the algebraic properties of a spacetime. We have also only considered a few very simple examples of solutions to the Einstein equations. Further work should provide us with invaluable intuition for the interpretation of the various algebraic classes. Moreover, we have not discussed here the higher-dimensional extension of the Newman-Penrose formalism for the Einstein equations, which has been the subject of much previous work concerning, for instance, problems of existence and stability of solutions \cite{Godazgar:2009fi,Durkee:2009nm,Durkee:2010xq,Durkee:2010qu,Durkee:2010ea,TaghaviChabert:2010bm, TaghaviChabert:2011ex, Godazgar:2011sn,Godazgar:2012zq,Dias:2013hn,deFreitas:2015pda}. Another interesting problem to investigate with our formalism is the use of curvature (and Cartan) invariants to characterise spacetimes; see \cite{MacCallum:2015zaa} for a brief introduction and \cite{D.D.McNutt:2017wkr,McNutt:2017paq,Coley:2017woz,Coley:2017vxb} for recent work on this topic.

To the obvious possible directions mentioned above, we add one further direction that we already alluded to in the introduction. This is the `double copy' between gauge theory and gravity, which appeared in the context of scattering amplitudes, and whose application to classical solutions is now under study. The existence of an analogy is, of course, natural from discussions such as the one in this paper, when comparing the classifications of the field strength tensor and the Weyl tensor. The point is, however, that there is a precise formulation of the double copy in this context. This is the subject of work in progress, and it was an important motivation for us to revisit the classification problem in this paper.

\acknowledgments
We thank Alan Coley, Christian Br\o nnum-Hansen, Mahdi Godazgar and Lionel Mason for useful discussions. RM is supported by a Royal Society University Research Fellowship. IN is supported by STFC studentship ST/N504051/1. DOC is an IPPP associate, and thanks the IPPP for on-going support as well as for hospitality during this work. He is supported in part by the Marie Curie FP7 grant 631370 and by the STFC consolidated grant ``Particle Physics at the Higgs Centre". 

\appendix

 \section{Multi-irrep spacetimes in the de Smet classification}\label{CompositeSpacetimes}
 
In section \ref{Section::deSmet}, it was shown that the de Smet classification is highly sensitive to the presence of a single little group irrep. What about when more than one irrep contributes to the Weyl tensor? Generically, this will lead to a $\textbf{4}$. For example, it can be seen from the discussion in section~\ref{Section::deSmet} that combining a $\textbf{22}$ or a $\textbf{\underline{22}}$ with a \textbf{1111} will always produce a \textbf{4}. Similarly, while de Smet classes are invariant under the interchange $k\leftrightarrow n$, combining any irrep with its $k\leftrightarrow n$ pair creates a $\textbf{4}$, if the two irreps are distinct. However, there are two cases when more than one irrep is present and the spacetime is still special in the de Smet classification:
 
 \begin{itemize}
 	\item Absence of any $\psi^{(i)}$
 	
 	The Weyl polynomials of all four irreps of dimension 3 or less contain a factor $\left[ n\cdot\xi,k\cdot\xi \right]$. This means that when only irreps of dimension 3 or less are present in the spacetime, they will in general form a $\textbf{22}$. However, if $\chi^{(1)}$, $\chi^{(2)}$ and $\chi^{(3)}$ are present and all directly proportional to each other, they can form into a $\textbf{211}$. This works as follows. Let the $\chi^{(i)}$ factorise as
 	\begin{equation}
 	\chi^{(1)}_{ab}= X\,  \theta_{(a} \kappa_{b)}, \hspace{10pt}		\chi^{(2)}_{ab}= Y\,  \theta_{(a} \kappa_{b)},\hspace{10pt}		\chi^{(3)}_{ab}= Z\,  \theta_{(a} \kappa_{b)}.
 	\end{equation}
 	Now the Weyl polynomial is of the form
 	\begin{equation}
 	\begin{split}
 	\mathcal{W}=\;-3 \left[ \sn\sdot\xi, \sk\sdot\xi \right]&
 	\big\{ 
 	X \, \left[\sn\sdot\xi ,\theta\right]\left[\sn\sdot\xi,\kappa\right]
 	+Y\, \left[\sn\sdot\xi ,\theta \right]\left[\sk\sdot\xi,\kappa\right]
 	\\
 	&
 	+Y\,\left[\sn\sdot\xi,\kappa\right] \left[\sk\sdot\xi,\theta\right]
 	+Z \left[\sk\sdot\xi,\theta\right]\left[\sk\sdot\xi,\kappa\right]
 	\big\},
 	\end{split}
 	\end{equation}
 	which factorizes into a \textbf{211} if $X\,Z=Y^2$:
 	\begin{equation}
 	\mathcal{W} = -3 \left[ \sn\sdot\xi, \sk\sdot\xi \right]\left(X \, \left[\sn\sdot\xi,\theta\right]+Y\, \left[\sk\sdot\xi,\theta\right]\right) \left(\left[\sn\sdot\xi,\kappa\right]+\frac{Y}{X}\,\left[\sk\sdot\xi,\kappa\right]\right).
 	\end{equation}
 	In other words, if the three vectors $\underline{\chi_1}$, $\underline{\chi_2}$ and $\underline{\chi_3}$ all point in the same direction with relative magnitudes satisfying $|\underline{\chi_1}|\, |\underline{\chi_3}| =|\underline{\chi_2}|^2$ then a special \textbf{211} composite spacetime is formed.
 	
 	\item $ \textbf{211} \, +\, \textbf{1111} $
 	
 	If the Weyl tensor contains only non-zero $\psi^{(4)}$ and $\chi^{(3)}$ terms (or $\psi^{(0)}$ and $\chi^{(1)}$), it is possible for these to form a de Smet $\textbf{31}$ or $\textbf{211}$. Let us define
 	\begin{equation}
\psi^{(4)}_{abcd}=\alpha^{(4)}_{(a}\,\beta^{(4)}_b\,\gamma^{(4)}_c\,\delta^{(4)}_{d)},\qquad \chi^{(3)}_{ab}=\theta^{(3)}_{(a}\,\kappa^{(3)}_{b)}.
 	\end{equation} Now, if one direction is the same, for example $\theta^{(3)}\propto \alpha^{(4)}$, then the Weyl polynomial forms a $\textbf{31}$, 
 	\begin{equation}
 	\begin{split}
 	\mathcal{W}=\left[\sk\sdot\xi,\alpha^{(4)}\right]\bigg\{&\;\left[\sk\sdot\xi,\beta^{(4)}\right]\left[\sk\sdot\xi ,\gamma^{(4)}\right]\left[\sk\sdot\xi,\delta^{(4)}\right] 
	\\&+ \frac{|\theta^{(3)}|}{|\alpha^{(4)}|}\left[\sk\sdot\xi,\kappa^{(3)}\right] \left[n\sdot\xi,\sk\sdot\xi\right] \bigg\},
 	\end{split}
 	\end{equation}
 	while if two directions are shared such that  $\theta^{(3)}\propto \alpha^{(4)}$ and  $\kappa^{(3)}\propto \beta^{(4)}$ then the Weyl polynomial remains a $\textbf{211}$,
 	\begin{equation}
 	\begin{split}
 	\mathcal{W}=\left[\sk\sdot\xi,\alpha^{(4)}\right]\left[\sk\sdot\xi,\beta^{(4)}\right]\bigg\{&\;\left[\sk\sdot\xi ,\gamma^{(4)}\right]\left[\sk\sdot\xi,\delta^{(4)}\right] \\&+ \frac{|\theta^{(3)}|}{|\alpha^{(4)}|} \frac{|\kappa^{(3)}|}{|\beta^{(4)}|}\, \left[\sn\sdot\xi,\sk\sdot\xi\right] \bigg\}.
 	\end{split}
 	\end{equation}
 
 In contrast, if $\psi^{(4)}$ is of the special de Smet form $\textbf{\underline{11}\,\underline{11}}$ and shares a direction with $\chi^{(3)}$, then the spacetime is always a $\textbf{211}$: the reality conditions prevent us from constructing a $\textbf{31}$. This is because the reality conditions on a $\psi^{(4)}$ of the form
 \begin{equation}
 \psi^{(4)}_{abcd}=\alpha_{(a}\,\beta_b\,\alpha_c\,\beta_{d)}
 \end{equation}
 are 
 \begin{equation}
 \alpha_1 \, \beta_1 = \pm (\alpha_2 \,\beta_2)^*,\hspace{10pt} \alpha_1\, \beta_2 + \alpha_2 \, \beta_1 = \mp \left(\alpha_1\, \beta_2 + \alpha_2 \, \beta_1 \right)^*,
 \end{equation}
 requiring a $\beta$ that looks like
 \begin{equation}\label{RealityConditionsForPsi1111}
 \beta=\begin{pmatrix}
 1 \\ -\alpha_1^*\,/\,\alpha_2^*
 \end{pmatrix}\,\beta_1, \hspace{10pt} \beta_1^*=\mp\, \frac{\alpha_2}{\alpha_2^*} \,\beta_1.
 \end{equation}
 The reality conditions for $\chi_3$ of the form $	\chi^{(3)}_{ab}=\theta_{(a}\,\kappa_{d)}
 $ are very similar:
 \begin{equation}
 \theta_1 \,\kappa_1 =  (\theta_2 \,\kappa_2)^*,\hspace{10pt} \theta_1\, \kappa_2 + \theta_2 \,\kappa_1 = - \left(\theta_1\, \kappa_2 + \theta_2 \, \kappa_1 \right)^*,
 \end{equation}
 with solution
 \begin{equation}\label{RealityConditionsForChi}
 \kappa=\begin{pmatrix}
 1 \\ -\theta_1^*\,/\,\theta_2^*
 \end{pmatrix}\,\kappa_1, \hspace{10pt}\kappa_1^*=-\, \frac{\theta_2}{\theta_2^*} \,\kappa_1.
 \end{equation}
 Therefore, if $\psi^{(4)}$ and $\chi_3$ share a direction such that $\alpha \propto \theta$, then it can be read off from equations \eqref{RealityConditionsForPsi1111} and \eqref{RealityConditionsForChi} that $\beta$ and $\kappa$ are proportional.  
  	
  \end{itemize}
 
\begin{figure}[h]
 	\centering
 	\includegraphics[width=.9\linewidth]{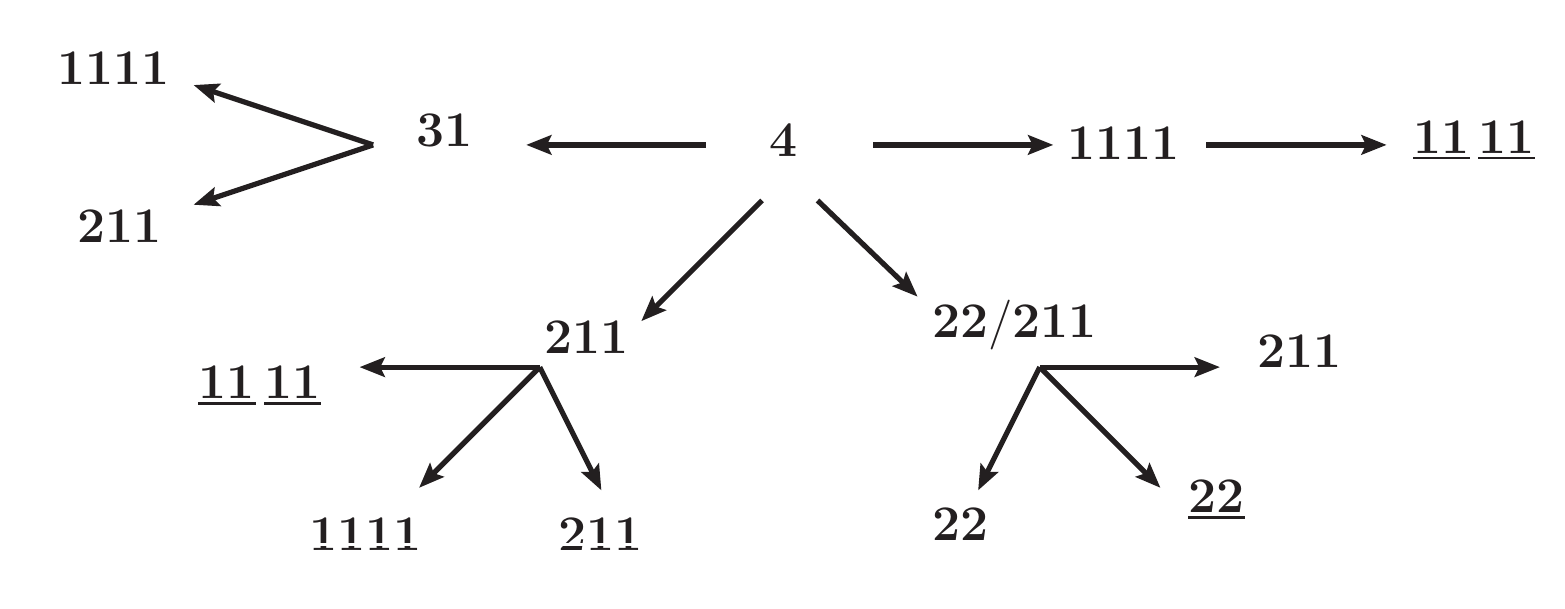}
 	\caption{There are 4 ways that the de Smet classes can become more specialised. Going clockwise from the top: a type N solution can become more special when its eigenvalues are equal. A spacetime containing more than one irrep of dimension 1 or 3 can be a $\textbf{22}$ or a \textbf{211} if the dimension 3 irreps form a perfect square. A \textbf{211} spacetime can also be formed using the irreps dimension 5 irreps, and a \textbf{31} spacetime always is.}
 	\label{figure::ImprovedDeSmetClassification}
 \end{figure}

 These are the only ways that a de Smet class can be built - every other combination results in a $\textbf{4}$. Figure \ref{figure::deSmetClassification} is therefore misleading, since it implies that each class can be reduced to another wholly contained within it. For example, figure \ref{figure::deSmetClassification} implies that de Smet \textbf{1111}s are a subset of \textbf{211}s. This is not always the case: a spacetime with only $\chi^{(3)}$ non-zero has no overlap with a spacetime which has only $\psi^{(0)}$ non-zero. An attempt to depict this limited specialisation of de Smet classes more accurately has been made in figure \ref{figure::ImprovedDeSmetClassification} as a contrast to figure \ref{figure::deSmetClassification}.
 
	\bibliographystyle{JHEP}
	\bibliography{references}

\end{document}